\documentclass[12pt]{article}
\usepackage[totalwidth=470truept,totalheight=636truept]{geometry}
\usepackage{latexsym,graphicx,amsfonts,amssymb,amsmath,color,dsfont,cancel,tikz,multirow,slashed}
\usepackage[hypertexnames=false,hidelinks]{hyperref}
\usepackage{diagbox}
\usepackage{slashbox}
\usepackage[export]{adjustbox}
\usepackage[format=hang,font=small,textfont=it]{caption}

\newcommand{\AB}{\includegraphics[width=0.5in]{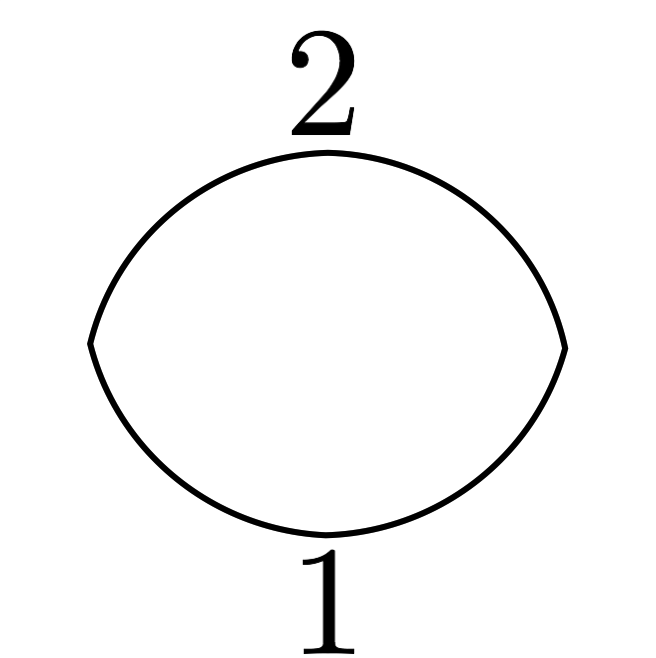}}
\newcommand{\ApBp}{\includegraphics[width=0.5in]{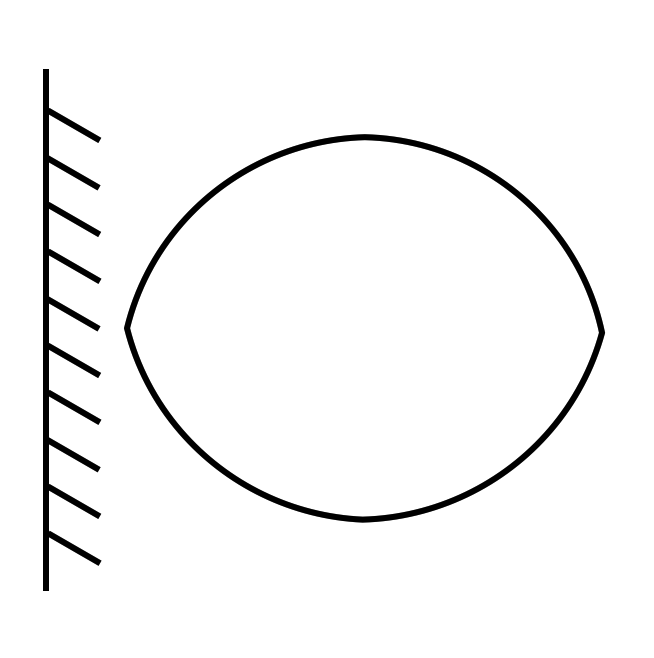}}
\newcommand{\ApB}{\includegraphics[width=0.5in]{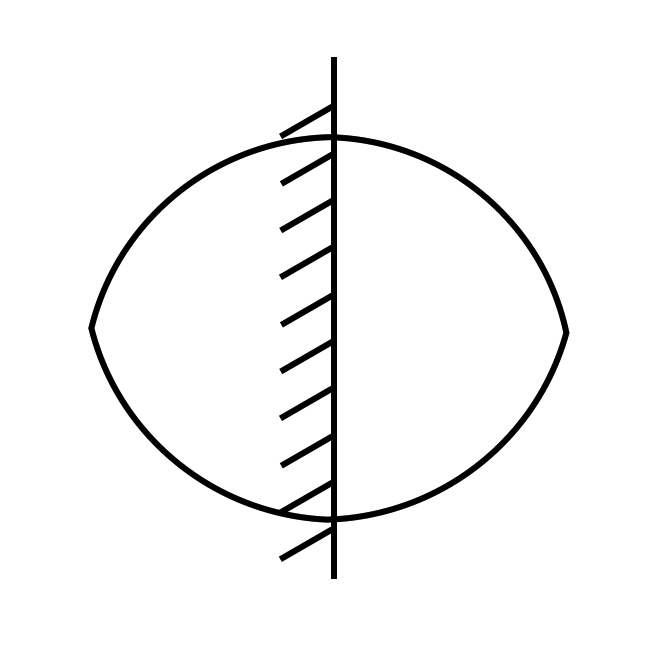}}
\newcommand{\ABp}{\includegraphics[width=0.5in]{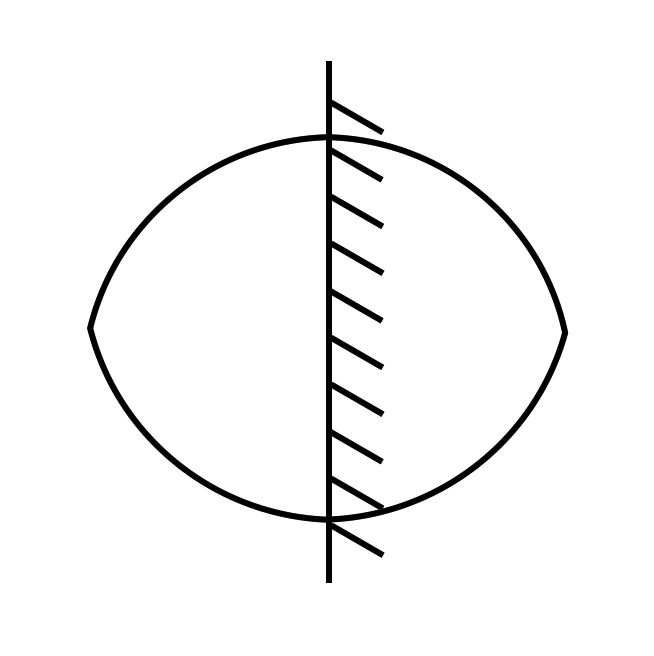}}

\newcommand{\ABC}{\includegraphics[width=0.48in]{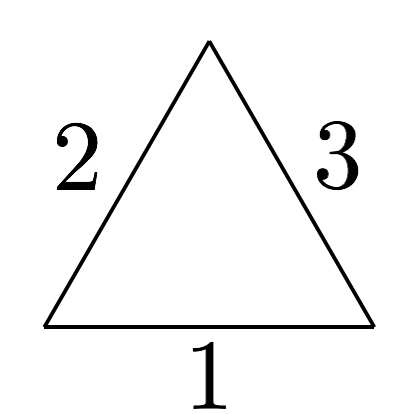}}
\newcommand{\ApBpCp}{\includegraphics[width=0.48in]{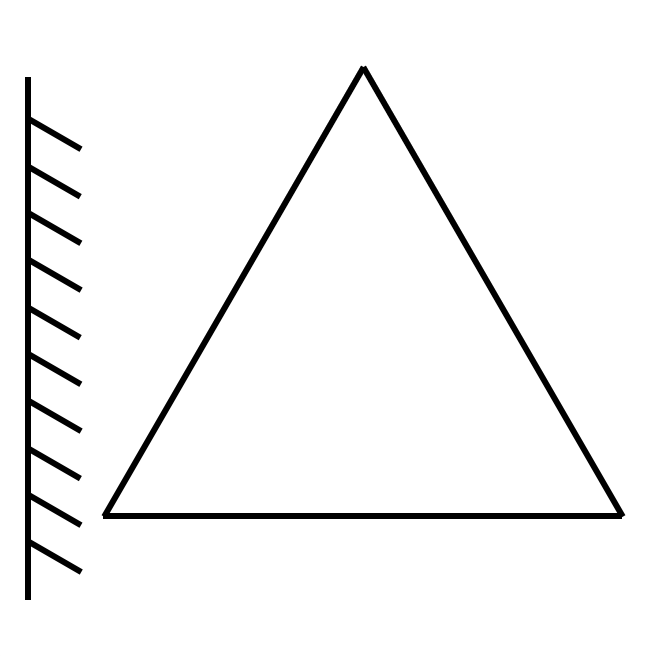}}
\newcommand{\ApBC}{\includegraphics[width=0.48in]{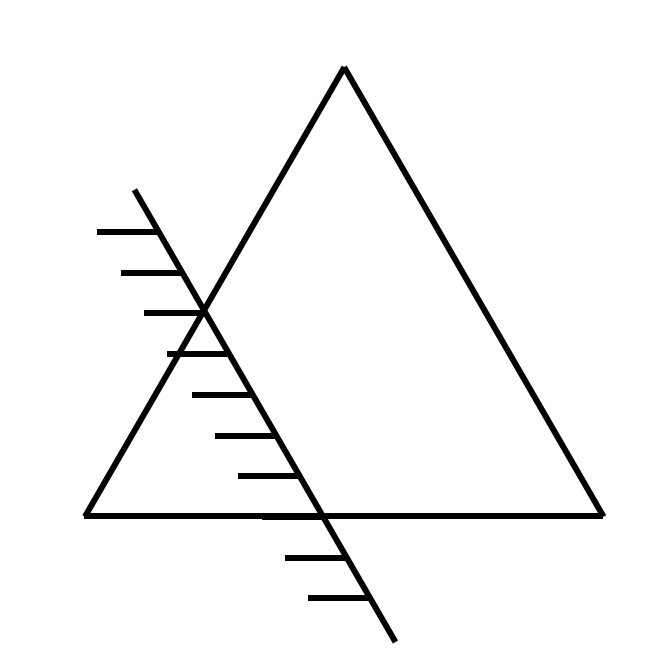}}
\newcommand{\ABpC}{\includegraphics[width=0.48in]{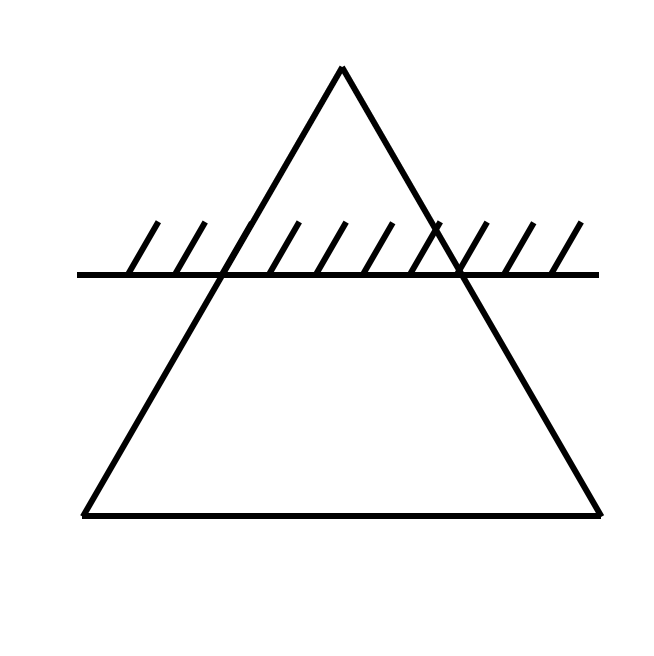}}
\newcommand{\ABCp}{\includegraphics[width=0.48in]{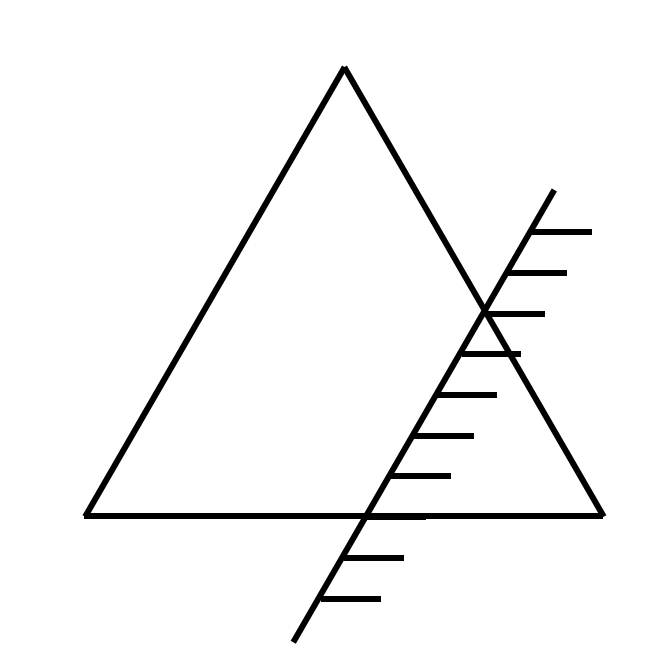}}
\newcommand{\ABpCp}{\includegraphics[width=0.48in]{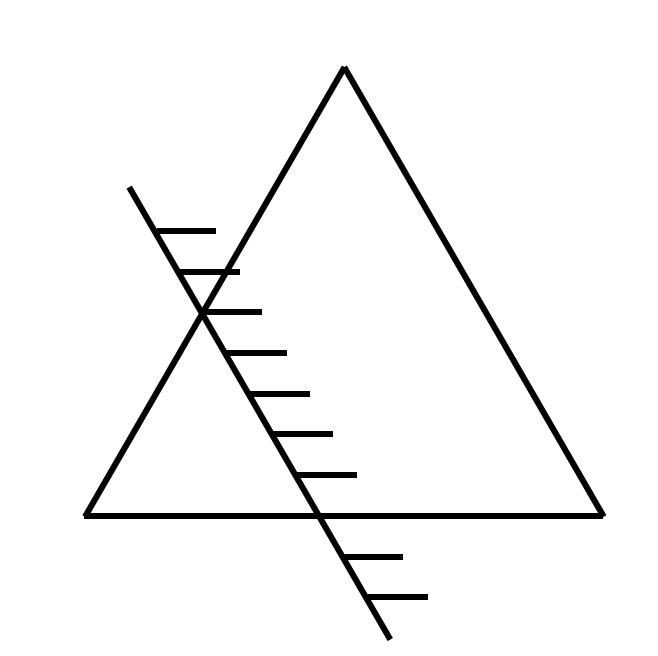}}
\newcommand{\ApBCp}{\includegraphics[width=0.48in]{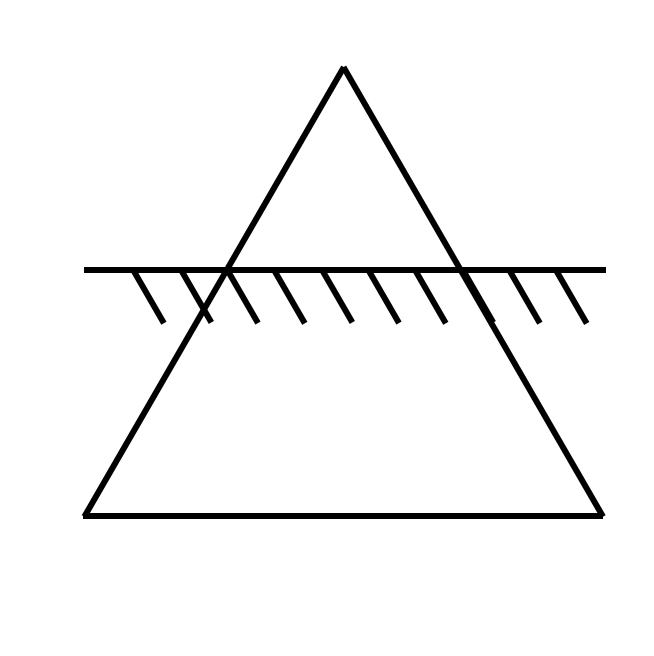}}
\newcommand{\ApBpC}{\includegraphics[width=0.48in]{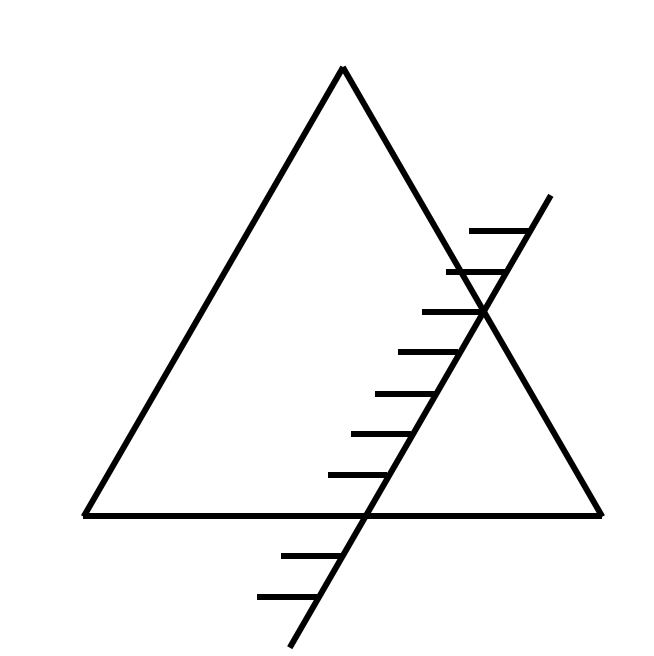}}

\DeclareMathAlphabet{\mymathbb}{U}{BOONDOX-ds}{m}{n}
\def\theequation{\arabic{section}.\arabic{equation}}
\newcommand{\sect}[1]{\setcounter{equation}{0}\section{#1}}
\renewcommand{\theequation}{\thesection.\arabic{equation}}
\global\arraycolsep=1truept
\renewcommand{\theequation}{\arabic{section}.\arabic{equation}}

\hypersetup{colorlinks=true,
    linkcolor=black,
    filecolor=black, 
    citecolor=black,
    urlcolor=blue}

\RequirePackage{booktabs}

\begin{document}

\null

\vskip1truecm

\begin{center}
{\LARGE \textbf{One-Loop Integrals}}
\vskip.8truecm
{\LARGE \textbf{for Purely Virtual Particles}}

\vskip1truecm

\textsl{{\large Aurora Melis$^{ \, \small{a}, \, 1}$ and Marco Piva$^{\, \small{b}, \, 2}$}}

\vskip .1truecm

{$^1$\textit{Dipartimento di Matematica e Fisica and INFN, Universit\`a di Roma Tre,\\ Via della Vasca Navale 84, Rome, Italy}}

\vspace{0.2cm}

{$^2$\textit{Faculty of Physics, University of Warsaw,\\ Pasteura 5, 02-093 Warsaw, Poland}}

\vspace{0.2cm}

$^{a}$aurora.melis@roma3.infn.it \ \ $^{\small{b}}$mpiva@fuw.edu.pl
\vskip1truecm

\textbf{Abstract}
\end{center}
Quantum field theories with purely virtual particles, or fakeons, require suitable modifications in one-loop integrals. We provide the expressions for the modified scalar integrals in the case of the bubble, triangle and box diagrams. The new functions are defined by means of their difference with the `t Hooft-Veltman scalar integrals. The modifications do not affect the derivation of the Passarino-Veltman reduction and one-loop integrals with nontrivial numerators can be decomposed in the same fashion. Therefore, the new functions can be directly used to study the phenomenology of any models with standard particles and fakeons. We compare our results with standard amplitudes and show that the largest differences are often localized in relatively small energy ranges and are characterized by additional nonanalyticities. Finally, we give explicit examples in the context of a toy model, where cross sections and decay widths of standard particles are modified by the presence of fakeons.

\vfill\eject

\section{Introduction}
Purely virtual particles, or fakeons, are a new type of degrees of freedom that can mediate interactions without appearing as external, on-shell states. Thanks to this property, they contribute to renormalization as standard particles, while they modify the behavior of observables, such as cross sections and decay rates. The consistency of this approach is guaranteed by a new quantization prescription~\cite{Anselmi:2017yux,Anselmi:2017ygm}, which allows us to project the fakeons away from the physical spectrum and preserves unitarity in the projected Fock subspace. The main application of purely virtual particles is quantum gravity~\cite{Anselmi:2018ibi,Anselmi:2018tmf}. Indeed, fakeons are able to reconcile renormalizability and unitarity, which has been the main issue in formulating a quantum field theory of gravitational interactions for a long time.

However, the prescription is rather general, and it can be applied to any quantum field theory, as long as the fakeons are massive and nontachyonic. Therefore, it is compulsory to study the presence of fakeons in the context of particle physics. The reason for suggesting this approach is twofold. On the one hand, as observed in \cite{Anselmi:2018yct}, there might still be room for some of the particles in the Standard Model (SM) to be purely virtual and this hypothesis deserves to be checked against present experimental data. On the other hand, thanks to their purely virtual nature, fakeons are prevented from being directly detected. It follows that if the new physics introduced in Beyond the Standard Model (BSM) models is (re)formulated to be purely virtual it can evade many experimental constraints which invalidate the most conventional models. Thus, phenomenology and model building should be rethought in light of the potential presence of purely virtual particles. An example of this is shown in~\cite{Anselmi:2021icc,Anselmi:2021chp} .

In high-energy physics, the study of cross sections and amplitudes at one loop can be eased thanks to the `t Hooft-Veltman (TV) functions~\cite{tHooft:1978jhc} and the Passarino-Veltman reduction~\cite{Passarino:1978jh}. The former are the one-loop integrals for scalar theories, whose explicit expressions have been obtained in~\cite{tHooft:1978jhc} and their analytic properties studied by several authors~\cite{Karplus:1958zz,Frederiksen:1971sj,Rechenberg:1972rq,Frederiksen:1973sk, Frederiksen:1973sz, Denner:1991kt}. The latter is a procedure which allows to write any one-loop integral in terms of the scalar ones. These results, combined with the usage of softwares, such as {\tt FeynCalc}, {\tt FormCalc} and {\tt LoopTools}~\cite{Kublbeck:1990xc,Hahn:1999mt}, help to systematize the one-loop calculations in the SM and BSM theories, which are rather cumbersome, given the amount of diagrams that needs to be computed. Therefore, if we want to include fakeons in the picture, it is necessary to introduce modified TV-functions that take into account the new prescription and its effects on one-loop integrals. 

Recently, a new way of introducing fakeons has been formulated~\cite{Anselmi:2021hab}. Among other things, it provides the \emph{threshold decomposition}, which allows to prove unitarity in a more direct way by splitting the usual optical theorem into a set of identities that hold independently threshold by threshold. 

In this paper we make use of the threshold decomposition to derive the modified TV-functions and study their behavior compared with the standard ones. In particular, we compute the new TV-functions for the bubble, triangle and box diagrams. The new functions reduce to the standard ones when fakeons are absent. Moreover, the Passarino-Veltman reduction can be applied to the new functions with no obstruction. 

A crucial difference between the usual TV-functions and the modified ones is that the latter present additional nonanalyticities, such as discontinuities and points of non-differentiability. From the phenomenological point of view, it is remarkable that in a scattering cross section these features appear in rather localized ranges of the center-of-mass energy and their largest magnitude are often on the critical points, where such differences turn on. This allows to confine the domain of the center-of-mass energy where we might expect to experimentally detect new physical effects. In order to illustrate these aspects of theories with fakeons, we give some explicit examples by using a scalar toy model, which includes all the ingredients to appreciate the differences with standard cases. Some effects have also been studied in the context of the inert-doublet model~\cite{Anselmi:2021icc} and the muon-$g-2$ anomaly~\cite{Anselmi:2021chp}.

The paper is organized as follows. In~\autoref{sec:prescription} we review the fakeon prescription and the threshold decomposition. In~\autoref{sec:PVmod} we derive the modified TV-functions for the bubble, triangle and box diagrams, describe their features and point out the configurations in the external momenta where the differences with the standard functions appear. In~\autoref{sec:toy} we present a toy model and show the modifications in the cross sections induced by the presence of fakeons, using the functions obtained in~\autoref{sec:PVmod}. Section~\ref{sec:conclusions} contains our conclusions.

We use the Minkowksi metric $\eta_{\mu\nu}=\text{diag}(1,-1,-1,-1)$.

\section{Threshold decomposition and fakeon prescription}
\label{sec:prescription}
In this section we review unitarity in terms of the \emph{threshold decomposition} introduced in~\cite{Anselmi:2021hab}, which is used to formulate the optical theorem as a set of independent identities that hold threshold by threshold. Moreover, such formulation manifestly shows that the optical theorem holds when the fakeon prescription is adopted.

\begin{figure}[t]
    \centering
    \includegraphics[scale=0.17]{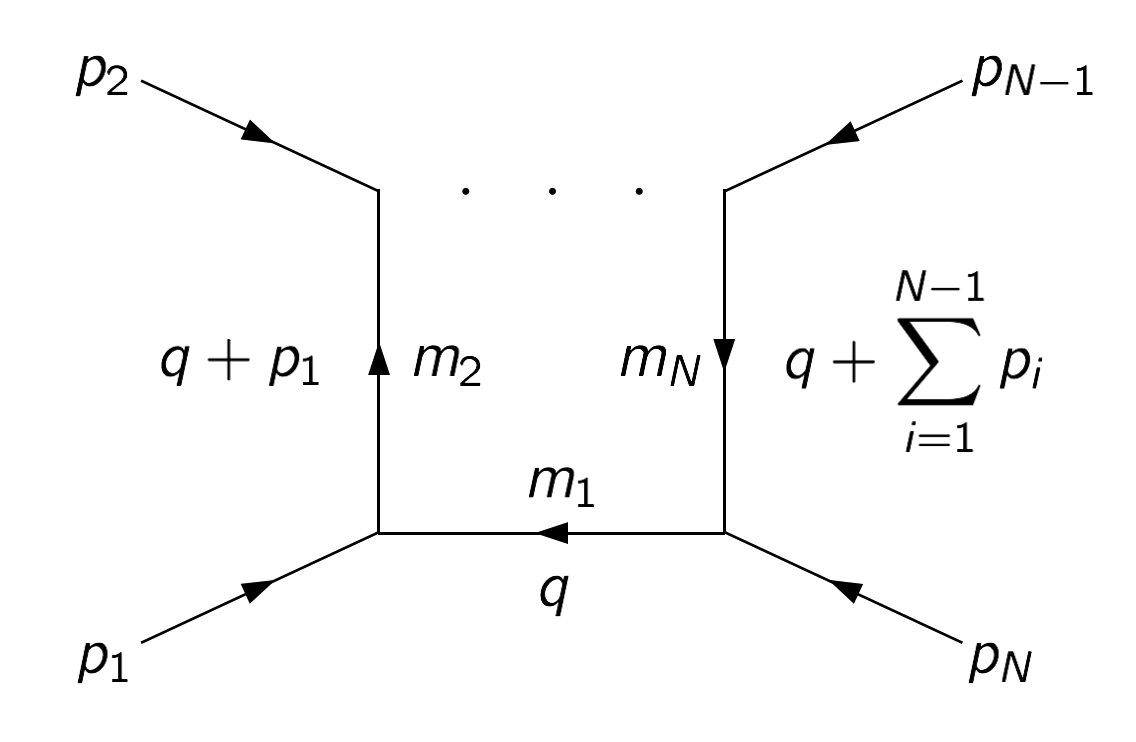}
    \caption{Momentum conventions adopted in this article.}
    \label{fig:Mom_Conv}
\end{figure}

The unitarity condition $SS^{\dagger}=\mathds{1}$ on the $S$-matrix can be written in the form of the optical theorem
\begin{equation}
    -i\left(T-T^{\dagger}\right)=TT^{\dagger}, \qquad S=\mathds{1}+iT.
\end{equation}
This equation can be decomposed diagrammatically in a set of identities, known as \emph{cutting equations}~\cite{Cutkosky:1960sp}, which for a single diagram $G$ reads
\begin{equation}\label{eq:cutting}
    G+G^*+\sum_{\text{cuts}}G_c=0,
\end{equation}
where $G_c$ are the \emph{cut diagrams}. Each $G_c$ is obtained from $G$ by cutting it with a continuous line in all possible way using the \emph{cut propagator}
\begin{equation}
    \frac{1}{q^2-m^2+i\epsilon}\rightarrow -i(2 \pi)\theta(\pm q^0)\delta\left(q^2-m^2\right),
\end{equation}
where the sign in the $\theta$-function depends on the direction of the energy flow in the diagram.  Diagrammatically, this is represented by a shadowed region on the right (left) side of a cut if the energy flows towards the cut from left (right) to right (left). Vertices in the shadowed region are complex conjugate. The sum on the right-hand side of~\eqref{eq:cutting} runs over all possible cut diagrams that can be built.
A general scalar one-loop diagram $G_{ N}$ with $N$ external legs is
\begin{equation}
G_{N}=\int \frac{\mathrm{d}^{D}q}{(2\pi )^{D}}\prod\limits_{a=1}^{N}\frac{1}{%
(q+k_{a})^{2}-m_{a}^{2}+i\epsilon_a}=\int \frac{\mathrm{d}^{D-1}\boldsymbol{q}}{(2\pi )^{D-1}}\left(
\prod\limits_{a=1}^{N}\frac{1}{2\omega _{a}}\right) G_{N}^{s},
\end{equation}
where the \emph{skeleton diagram} $G^s_{N}$ is defined as
\begin{equation}
G_{N}^{s}=\int \frac{\mathrm{d}%
q^{0}}{2\pi }\prod\limits_{a=1}^{{N}}\frac{2\omega _{a}}{(q^{0}+k^0_{a})^{2}-%
\omega _{a}^{2}+i\epsilon_a},  \qquad \omega_a=\sqrt{(\boldsymbol{q}-\boldsymbol{k}_a)^2+m^2_a}.
\end{equation}
The definition for $L$-loop diagrams is straightforward. To write the expressions in a symmetric way, a momentum $q+k_a$ is associated to each internal leg, although it is redundant. In the next sections we label the internal and external momenta according to the {\tt LoopTools}~\cite{Hahn:1999mt} conventions as in Figure \ref{fig:Mom_Conv}, i.e. we set $k_1=0$ and $k_N=\sum_{i=1}^{N-1}p_i$. Moreover, we simplify the computations by choosing the most convenient Lorentz frame, depending on the diagram we are interested in (see~\autoref{appcomp}). 

A diagram $G_{N}$ can be seen as a function of complex external momenta which typically presents nonanalyticities, such as branch cuts. Those cuts are associated with thresholds of the form 
\begin{equation}\label{eq:threshold}
    K^2\geq \left(\sum_{i\in A}m_i-\lambda\sum_{i\in B}m_i\right)^2,
\end{equation}
where $K$ is a combination of external momenta, $A$ and $B$ are two different subsets of internal legs, and $\lambda=0,1$. The thresholds with $\lambda=0$ are called \emph{physical} and are associated with the production of physical particles, namely they represent the minimum value for the invariant $K^2$ such that the particles in $A$ become real. These thresholds are those that appear in the optical theorem (see below). On the other hand, thresholds with $\lambda=1$ are pathological, since they are associated with instabilities. They are called \emph{pseudo-thresholds} and might appear in theories where different prescriptions for the propagators are adopted in a inconsistent way. However, in the case where the fakeon prescription and the Feynman one are used in the same theory this problem does not occur~\cite{Anselmi:2020tqo}.

Finally, there are other nonanalyticities associated to thresholds, different from~\eqref{eq:threshold}. They are called \emph{anomalous thresholds}~\cite{Cutkosky:anomalous} and they are not associated to the physical production of particles, although they might appear in physical amplitudes (see for example~\cite{Passarino:2018wix}).

\subsection{Threshold decomposition and spectral identities}
The skeleton diagrams satisfy the \emph{spectral optical theorem}~\cite{Anselmi:2021hab}
\begin{equation}\label{eq:spectraloptical}
G^{s}+(G^s)^*+\sum_{\text{cuts}}G_{c}^{s}=0.
\end{equation}
At a first look, it may seem that~\eqref{eq:spectraloptical} is just~\eqref{eq:cutting} where the diagrams are substituted with their skeleton version. What makes~\eqref{eq:spectraloptical} convenient is that the skeleton diagrams can be further decomposed into sum of terms which are associated to single physical thresholds, leading to the \emph{spectral identities}. These identities are derived once the integrals over the loop energies in the skeleton diagrams are performed by means of the residue theorem. In this way unitarity is proved without the need of integrating over the space components of loop momenta.

Labelling $e_a\equiv k^0_a$, the decomposition of a skeleton diagrams is given in terms of the following quantities
\begin{equation}
\mathcal{P}^{ab}=\mathcal{P}\frac{1}{e_{a}-e_{b}-\omega
_{a}-\omega_{b}}, \quad \mathcal{Q}^{ab}=\mathcal{P}\frac{2\omega _{b}}{(e_{a}-e_{b}-\omega
_{a})^2-\omega^2_{b}},\quad
\Delta ^{ab}=\pi \delta (e_{a}-e_{b}-\omega _{a}-\omega_{b}),
\label{eq:PQDdef}
\end{equation}
where $\mathcal{P}$ is the Cauchy principal value. Moreover, we define the quantity $\mathcal{P}_n$, which is a sum of product of $n-1$ different $\mathcal{P}^{ab}$. In this paper we do not need the explicit form of those terms and we only write them symbolically. Their expressions in terms of $\mathcal{P}^{ab}$'s can be found in~\cite{Anselmi:2021hab}.

The function $\Delta^{ab}$ is related to the diagram where the internal lines $a$ and $b$ are cut and the shadowed region is in the direction where the momentum $p_a$ enters. Accordingly, $\Delta^{ba}$ is associated with the cut diagram with the opposite shadowed region. As a rule of thumb, each $\Delta^{ab}$ is associated to the threshold condition $p^2_{a}>(m_a+m_b)^2$ if $e_a>e_b$ or $p^2_{b}>(m_a+m_b)^2$ if $e_a<e_b$.

In this paper we are interested in the the bubble, triangle and box diagrams. Therefore, in this section we report their threshold decomposition, which is derived as follows: (i) perform the integral over the loop energies by means of the residue theorem, (ii) write the skeleton diagrams as a sum of terms where only sum of frequencies $\omega_i$ appear in the denominators, (iii) use the relation
\begin{equation}\label{eq:PPDelta}
    \frac{i}{x+i\epsilon}=\mathcal{P}\frac{i}{x}+\pi\delta(x)
\end{equation}
everywhere, (iv) rewrite the result exclusively in terms of the quantities in~\eqref{eq:PQDdef}.

For example, consider the skeleton bubble diagram
\begin{equation}
    B^s\equiv G_2^s=\int\frac{\text{d}q^0}{2\pi}\frac{2 \omega_1}{(q^0+k_1^0)^2-\omega_1^2+i\epsilon_1}\frac{2 \omega_2}{(q^0+k_2^0)^2-\omega_2^2+i\epsilon_2}.
\end{equation}
After following step (i) and rescaling the epsilons we obtain
\begin{eqnarray}\label{eq:skelbub}
    B^s=&&-\frac{i}{e_1-e_2-\omega_1-\omega_2+i\epsilon_1+i\epsilon_2}+\frac{i}{e_1-e_2+\omega_1-\omega_2-i\epsilon_1+i\epsilon_2} \nonumber \\
    &&-\frac{i}{e_2-e_1-\omega_2-\omega_1+i\epsilon_1+i\epsilon_2}+\frac{i}{e_2-e_1+\omega_2-\omega_1+i\epsilon_1-i\epsilon_2}.
\end{eqnarray}
The the second and the fourth terms are ill-defined distributions containing differences of frequencies and they should disappear. Indeed, they cancel out. Setting $\epsilon_1=\epsilon_2=\epsilon/2$ and using \eqref{eq:PPDelta}, the threshold decomposition for the bubble skeleton diagram reads
\begin{equation}\label{Bubdecomp}
B^{s}=-i\mathcal{P}_2-\Delta ^{12}-\Delta ^{21},
\end{equation}
where $\mathcal{P}_2=\mathcal{P}^{12}+\mathcal{P}^{21}$. The same procedure is applied to derive the threshold decomposition for any other diagram albeit with some additional caveats~\cite{Anselmi:2021hab}.
To obtain the spectral identities we need to compute also the cut skeleton diagrams in a similar way. Once they are decomposed, it is useful to put them in a table, together with the original diagram and its complex conjugate, as shown in~\autoref{tb1}.
\begin{table}[t]
\begin{center}
\setlength{\tabcolsep}{3pt}
\renewcommand{\arraystretch}{1.1}
\newcolumntype{P}[1]{>{\centering\arraybackslash}p{#1}}
\newcolumntype{M}[1]{>{\centering\arraybackslash}m{#1}}
 \begin{tabular}{|P{1.3cm}|M{1.2cm}|M{1.2cm}|M{1.2cm}|M{1.2cm}|}
 \hline
 \diagbox[innerwidth=1.3cm,height=1.5cm]{\small Terms}{\small Diag.} & \AB & \ApBp  & \ABp & \ApB\\ \hline 
$\mathcal{P}_{2}$ & $-i$ & $i$ & $0$ & $0$ \\ 
\hline
$\Delta ^{12}$ & $-1$ & $-1$ & $2$ & $0$ \\ \hline
$\Delta ^{21}$ & $-1$ & $-1$ & $0$ & $2$ \\ \hline
\end{tabular}
\end{center}
\caption{Threshold decomposition of the bubble diagram.}
\label{tb1}

\end{table}
In the columns we place the diagrams, while in the rows we place the coefficients that multiply the terms of the threshold decomposition. The sum of each row cancels independently. These are the spectral identities. 
From this decomposition we see that the usual optical theorem is actually the sum of identities that hold independently. Note that each row (besides the first one) is associated to a different threshold. This can be better appreciated in the decomposition of the skeleton triangle diagram, which reads~\cite{Anselmi:2021hab}
\begin{equation}
C^{s}=-i\mathcal{P}_{\text{3}}+\sum_{\text{perms}}\left[-\Delta ^{ab}%
\mathcal{Q}^{ac}+\frac{i}{2}\Delta ^{ab}(\Delta
^{ac}+\Delta ^{cb})\right].  \label{Tdecomp}
\end{equation}%

\begin{table}[t]
\begin{center}
\setlength{\tabcolsep}{3pt}
\renewcommand{\arraystretch}{1.1}
\newcolumntype{M}[1]{>{\centering\arraybackslash}m{#1}}
\newcolumntype{P}[1]{>{\centering\arraybackslash}p{#1}}
\begin{tabular}{|P{1.5cm}|M{1.5cm}|M{1.3cm}|M{1.3cm}|M{1.3cm}|M{1.3cm}|M{1.3cm}|M{1.3cm}|M{1.3cm}|M{1.3cm}|M{1.3cm}|}
\hline
\diagbox[innerwidth=1.5cm,height=1.5cm]{\small Terms}{\small Diag.}  & \ABC & \ApBpCp & \ABpCp & \ApBCp & \ApBpC
& \ApBC & \ABpC & \ABCp \\ \hline
$\mathcal{P}_{3}$ & $-i$ & $i$ & $0$ & $0$ & $0$ & $0$ & $0$ & $0$ \\ \hline
$\Delta^{12}\mathcal{Q}^{13}$ & $-1$ & $-1$ & $2$ & $0$ & $0$ & $0$ & $0$ & $0$ \\ \hline
$\Delta^{23}\mathcal{Q}^{21}$ & $-1$ & $-1$ & $0$ & $2$ & $0$ & $0$ & $0$ & $0$ \\ \hline
$\Delta^{31}\mathcal{Q}^{32}$ & $-1$ & $-1$ & $0$ & $0$ & $2$ & $0$ & $0$ & $0$ \\ \hline
$\Delta^{21}\mathcal{Q}^{23}$ & $-1$ & $-1$ & $0$ & $0$ & $0$ & $2$ & $0$ & $0$ \\ \hline
$\Delta^{32}\mathcal{Q}^{31}$ & $-1$ & $-1$ & $0$ & $0$ & $0$ & $0$ & $2$ & $0$ \\ \hline
$\Delta^{13}\mathcal{Q}^{12}$ & $-1$ & $-1$ & $0$ & $0$ & $0$ & $0$ & $0$ & $2$ \\ \hline
$\Delta^{12}\Delta^{13}$ & $i$ & $-i$ & $2i$ & $0$ & $0$ & $0$ & $0$ & $-2i$ \\ \hline
$\Delta^{23}\Delta^{21}$ & $i$ & $-i$ & $0$ & $2i$ & $0$ & $-2i$ & $0$ & $ 0 $ \\ \hline
$\Delta^{31}\Delta^{32}$ & $i$ & $-i$ & $0$ & $0$ & $2i$ & $0$ & $-2i$ & $ 0 $ \\ \hline
$\Delta^{21}\Delta^{31}$ & $i$ & $-i$ & $0$ & $0$ & $2i$ & $-2i$ & $0$ & $ 0 $ \\ \hline
$\Delta^{32}\Delta^{12}$ & $i$ & $-i$ & $2i$ & $0$ & $0$ & $0$ & $-2i$ & $0$ \\ \hline
$\Delta^{13}\Delta^{23}$ & $i$ & $-i$ & $0$ & $2i$ & $0$ & $0$ & $0$ & $-2i $ \\ \hline
\end{tabular}%
\end{center}
\caption{Threshold decomposition of the triangle diagram.}
\label{ts1}
\end{table} 
From~\eqref{Tdecomp} and its cut diagrams we obtain~\autoref{ts1}. The terms with a single $\Delta$ are associated with physical thresholds, while those with double $\Delta$ are associated to unphysical thresholds. Indeed, the former contribute to the optical theorem~\eqref{eq:cutting}, which states that the sum of all cut diagrams cancel with the diagram and its conjugate, while the latter do not. This can be seen from~\autoref{ts1}. In fact, the coefficients in the rows with a single $\Delta$ cancel between diagrams and cut diagrams. On the other hand, the coefficients in the rows with two $\Delta$ cancel between cut diagrams and between the diagram and its conjugate, independently. This shows explicitly that those thresholds do not contribute to the optical theorem. 

Finally, the threshold decomposition of the box diagram reads~\cite{Anselmi:2021hab}
\begin{equation}\label{Boxdecomp}
    D^s=-i\mathcal{P}_4+\sum_{\text{perms}}\left[-\frac{1}{2}\Delta^{ab}\mathcal{Q}^{ac}\mathcal{Q}^{ad} + \frac{i}{2}\Delta^{ab}(\Delta ^{ac}+\Delta ^{cb})\mathcal{Q}^{ad}+\frac{1}{6}\Delta^{ab}(\Delta ^{ac}\Delta ^{ad}+\Delta ^{cb}\Delta ^{db})\right].
\end{equation}
Again, the terms with more than one $\Delta$ are associated with unphysical thresholds and they do not contribute to the optical theorem.

In general, we can always write a diagram $G_{N}$ as
\begin{equation}
    G_{ N} = -i \mathcal{P}_{ N} + G_{N}^{\Delta},
\end{equation}
where $\mathcal{P}_{N}$ encodes the purely virtual content of the diagram, while $G_{N}^{\Delta}$ contains all the terms with at least one $\Delta$ and gives information about the the production of on-shell physical particles. This is of particular use in the case of the fakeon prescription.

\subsection{The fakeon prescription}
Purely virtual particles, or fakeons, are degrees of freedom of a new type. They cannot be produced on shell, but can mediate interactions as virtual particles. They are introduced by means of a different quantization prescription, called \emph{fakeon prescription}~\cite{Anselmi:2017yux,Anselmi:2017ygm}, which modifies the diagrams so the optical theorem is satisfied, once we restrict to the Fock subspace where the fakeons are not external lines. Roughly speaking, such modifications remove the parts of the amplitudes that contain information about the fakeons being on shell. As a result, every time a fakeon internal line is cut, the associated cut diagram is zero and loop effects cannot resuscitate the fakeons once they are removed from the possible external lines. This prevents those type of particles from being on shell at arbitrary scales, making them radically different from resonances or unstable particles, which in principle can be observed on shell.

In terms of the threshold decomposition, the fakeon prescription is systematically implemented as follows. First, evaluate the skeleton diagrams following the steps (i) and (ii) described in the previous subsection. Then, substitute~\eqref{eq:PPDelta} with 
\begin{equation}\label{eq:PPDeltafake}
    \frac{i}{x+i\epsilon}\rightarrow \mathcal{P}\frac{i}{x}+\tau\pi\delta(x),
\end{equation}
where $\tau=0$ if $x$ contains at least one fakeon frequency, and $\tau=1$ otherwise. Finally, to explicitly show that the optical theorem holds, express the result in terms of the quantities~\eqref{eq:PQDdef}. However, for the only purpose of evaluating the amplitudes, this final step is optional. The substitution~\eqref{eq:PPDeltafake} with $\tau=0$ amounts to remove all the $\Delta$'s that contain at least a fakeon frequency $\omega$, i.e. $\Delta^{ab}=0$ if the leg $a$ and/or $b$ is a fakeon.

In practice, the effects of the fakeon prescription on the diagrams are the following.

1) At the tree level, formula~\eqref{eq:PPDeltafake} reduces to choose the propagator
\begin{equation}\label{eq:FW}
\mathcal{P}\frac{i}{p^2-m^2}
\end{equation}
for each fakeon leg. Note that the propagator~\eqref{eq:FW} coincides with the \emph{Feynman-Wheeler} one~\cite{Wheeler:1945ps}, which leads to inconsistencies, such as violation of the locality of counterterms and instabilities~\cite{Anselmi:2020tqo}, when used inside loop diagrams.

2) At every loop order, instead of using the propagator~\eqref{eq:FW}, a diagram $G$ that involves a fakeon in the internal lines becomes
\begin{equation}
    G^{\prime}=G-\Delta_{\rm f} G,
\end{equation}
where $\Delta_{\rm f} G$ is given by the terms in the threshold decomposition of $G$ that are removed by setting $\tau$ to zero. 
The simplest example that shows that $G^{\prime}$ is not the diagram built with~\eqref{eq:FW} is the bubble diagram. In fact, if we use~\eqref{eq:FW} its imaginary part is nonvanishing, while it is set to zero (see below) when at least one fakeon is in the loop and the inconsistencies mentioned in point 1) do not occur, as explained in~\cite{Anselmi:2020tqo}.

In the case of a triangle diagram where the leg 1 is a fakeon, \autoref{ts1} reduces to~\autoref{ts2}.
\begin{table}[t]
\begin{center}
\setlength{\tabcolsep}{3pt}
\renewcommand{\arraystretch}{1.1}
\newcolumntype{M}[1]{>{\centering\arraybackslash}m{#1}}
\newcolumntype{P}[1]{>{\centering\arraybackslash}p{#1}}
 \begin{tabular}{|P{1.3cm}|M{1.2cm}|M{1.2cm}|M{1.2cm}|M{1.2cm}|M{1.2cm}|}
 \hline
 \diagbox[innerwidth=1.3cm,height=1.5cm]{\small Terms}{\small \,Diag.} & \ABC & \ApBpCp & \ApBCp & \ABpC \\ \hline
 $\mathcal{P}_{3}$ & $-i$ & $i$ & $0$ & $0$\\ \hline
 $\Delta^{23}\mathcal{Q}^{21}$ & $-1$ & $-1$ & $2$ & $0$ \\ \hline
 $\Delta^{32}\mathcal{Q}^{31}$ & $-1$ & $-1$ & $0$ & $2$ \\ \hline
 \end{tabular}
\end{center}
\caption{Threshold decomposition of the triangle diagram when particle 1 is a fakeon.}
\label{ts2}
\end{table}
Note that some cut diagrams (those where leg 1 is cut) are zero after the reduction and do not appear in the table.

From the definition of the fakeon prescription it is easy to see that unitarity still holds. Indeed, setting $\Delta$'s to zero removes entire rows from the tables of the threshold decomposition. Since the single rows sum to zero independently, we obtained reduced tables that give the optical theorem~\eqref{eq:cutting} once we project onto the subspace where fakeons are not external states. This is why we say that fakeons are purely virtual and cannot be produced on-shell. Therefore, the fakeon prescription must be supplemented with a projection, for consistency. Viceversa, a projection of the external states without a prescription that modifies loop diagrams is inconsistent, since the states that are projected away would be generated back by loop corrections.

Finally, fakeons have other two peculiar properties due to their quantization prescription: the microcausality violation and the peak uncertainty. Details about the former can be found in ref.s~\cite{Anselmi:2018tmf, Anselmi:2018bra}. For the purpose of this paper we only need to keep in mind that the masses of the fakeons must be large enough to avoid conflicts with macrocausality~\cite{Anselmi:2019nie}. 

The latter property is related to the fakeon width $\Gamma_{\text{f}}$ and it states that it is not possible to determine the shape of the fakeon peak with arbitrary precision. The reason is that the new prescription violates analyticity and does not allow for a resummation of the dressed propagator around the peak. Such a feature can be due to our missing knowledge about the nonperturbative sector of the theory, or it can be associated to a new physical indeterminacy. The details of these peculiar effects can be found in~\cite{Anselmi:2020lfx}. In this paper we concentrate on the indirect effects of fakeons on physical observables such as cross sections and decay rates of standard particles.

We stress that below every fakeon threshold there is no difference between the Feynman prescription and the fakeon one. In particular, many properties of standard quantum field theory hold there. For example, the Appelquist-Carrazzone decoupling theorem is still valid, since if we send all the fakeon masses to infinity, then every thresholds associated to their production would be moved to infinity as well. In that limit the accessible physics is always below every fakeon threshold and therefore is not modified by the fakeon prescription. In other words, the low-energy expansion of a theory with heavy fakeons or a theory with heavy standard particles instead of fakeons is the same.

A final remark concerns the question of stability. 
The presence of instabilities depends on the location of the poles in the energy complex plane of loop integrals, which determines whether differences of frequencies appear in the denominators. In this respect, it is important to recall that the first step of the fakeon prescription is to perform the integral over the loop energies by means of the residue theorem. Here the location of the poles is selected with the usual Feynman prescription. Only after this the $\Delta$'s are set to zero. Therefore, instabilities cannot appear in the expressions, since they do not in the standard Feynman integrals. Indeed, the cancellation shown in~\eqref{eq:skelbub} is valid for arbitrary diagrams.
This is of particular importance for the case of ghosts (which is not discussed in this paper but it is the main motivation for the fakeon prescription). In fact, when ghosts are present it is often stated that the theory presents instabilities. We want to stress that this happens when the anti-Feynman prescription is chosen for the ghosts, while choosing the Feynman one avoids those instabilities at the price of explicit violation of unitarity. The latter is then restored by proceeding with the fakeon prescription that provides a theory that is stable and unitary.

In the next section, we use the threshold decomposition to obtain a generalized expression for the TV-functions, that accounts for both the Feynman and the fakeon prescriptions.

\sect{Modified one-loop scalar integrals}\label{sec:PVmod}

In this section we derive how the one-loop scalar integrals get modified when fakeons appear in the internal legs of the bubble, triangle and box diagrams\footnote{Tadpole diagrams are not modified since they do not involve any threshold.}. These modifications are general and can be applied to any theory. Typical scattering processes in particle-physics experiments can be studied straightforwardly by implementing the modifications into softwares such as {\tt FeynCalc}, {\tt FormCalc} and {\tt LoopTools}. Natural applications are BSM models where the new physics can be purely virtual. However, there is room even for the standard model to contain fakeons~\cite{Anselmi:2018yct}. 

In order to obtain the modified TV-functions we integrate the terms in~\eqref{Bubdecomp},~\eqref{Tdecomp} and~\eqref{Boxdecomp} over spatial momenta. It is more convenient to compute the terms that need to be removed in the fakeon case, rather than the whole modified function. In this way we never have to compute the $\mathcal{P}$-terms, since the products of $\Delta$'s and $\mathcal{Q}$'s are enough. Moreover, all the ultraviolet divergences that are subtracted by means of renormalization are contained in the $\mathcal{P}$-terms. We assume the dimensional regularization for those terms, while we compute the other terms directly in four dimensions, since they do not need to be regularized.

Keeping in mind that the fakeon prescription amounts to remove all the $\Delta$'s associated to fakeon frequencies, it is practical to introduce 
\begin{eqnarray}
\Delta^{ab}_{\rm f}\equiv 
\begin{cases}
 0 &\quad\text{if $a \wedge b \in $ standard particles}\\
 \Delta^{ab} &\quad\text{if $a \lor b \in $ fake particles}.
\end{cases}
\end{eqnarray}
Therefore, we define the subtraction terms as follows
\begin{eqnarray}
\Delta_{\rm f} A_{0} &=& 0 ,\\[10pt]
\label{eq:DeltaB0}
\Delta_{\rm f} B_{0} &=&\int\hspace{-1.5mm}\frac{\mathrm{d}^3\boldsymbol{q}}{(2 \pi)^3}\sum_{\substack{\text{perms}}}\frac{i\Delta_{\rm f}^{ab}}{4\omega_a\omega_b},\\[10pt]
\label{eq:DeltaC0}
\Delta_{\rm f} C_{0} &=&\int\hspace{-1.5mm}\frac{\mathrm{d}^3\boldsymbol{q}}{(2 \pi)^3}\sum_{\substack{\text{perms} }} \frac{\Delta_{\rm f}^{ab}}{8\omega_a\omega_b\omega_c} \left[i%
\mathcal{Q}^{ac}+\frac{1}{2}(\Delta_{\rm f}
^{ac}+\Delta_{\rm f}^{cb})\right],\\[10pt]
\label{eq:DeltaD0}
\Delta_{\rm f} D_0 &=& 
\int\hspace{-1.5mm}\frac{\mathrm{d}^3\boldsymbol{q}}{(2 \pi)^3} \sum_{\substack{\text{perms}}}\frac{\Delta_{\rm f}^{ab}}{16\,\omega_a\omega_b\omega_c\omega_d}  \bigg[ \frac{i}{2}\mathcal{Q}^{ac}\mathcal{Q}^{ad} + \frac{1}{2}(\Delta_{\rm f}^{ac}+\Delta_{\rm f}^{cb})\mathcal{Q}^{ad} \big. \\
 \left. - \frac{i}{6}(\Delta_{\rm f}^{ac}\Delta_{\rm f}^{ad}+\Delta_{\rm f}^{cb}\Delta_{\rm f}^{db}) \right]\,,\nonumber \hspace{-10cm} 
\end{eqnarray}
where the sum is over the permutations of the indices of the internal legs. 
Thus, the modified TV-functions are obtained from the standard ones by means of the subtractions
\begin{eqnarray}
\label{eq:A0f}
A_0^{\text{f}}&\equiv&A_0,\\
\label{eq:B0f}
B_0^{\text{f}}&\equiv&B_0-\Delta_{\rm f} B_0,\\
\label{eq:C0f}
C_0^{\text{f}}&\equiv&C_0-\Delta_{\rm f} C_0,\\
\label{eq:D0f}
D_0^{\text{f}}&\equiv&D_0-\Delta_{\rm f} D_0.
\end{eqnarray}
In this way all the terms associated to thresholds that involve at least one fakeon are removed from the original functions. 
The new TV-functions defined above are general and include the standard ones as particular cases. Indeed, when fakeons are absent $\Delta_{\rm f} B_0, \Delta_{\rm f} C_0$ and $\Delta_{\rm f} D_0$ vanish and the modified functions coincide with the usual ones. In the rest of this section we focus on computing the terms in (\ref{eq:DeltaB0}-\ref{eq:DeltaD0}) and identify the configurations where they are nonzero. 

\subsection{$B^{\rm f}_0$ function}\label{sec:B0f_function}

We start with the simplest case of the one-loop scalar two-point function $B_0(p^2_1,m^2_1,m^2_2)$ associated with the bubble diagram. The $\Delta$-term is equal to the imaginary part of $B_0$ and gives the well-known result
\begin{equation}
\label{eq:D12term}
    \int\frac{\mathrm{d}^3\boldsymbol{q}}{(2 \pi)^3}\frac{ \Delta^{12}}{4\,\omega_1\omega_2}=\frac{\theta\left(p^2_1-(m_1+m_2)^2\right)\sqrt{\lambda\left(p^2_1,m^2_1,m^2_2\right)}}{16\,\pi\,p^2_1},
\end{equation}
where $\theta(x)$ is the Heaviside step function\footnote{In case of multiple conditions we adopt the compact notation $\theta(x,\cdots,y)\equiv \theta(x)\cdots\theta(y)$.} and  $\lambda$ is the  K\"allén function associated with the area of the triangle of sides $x,y,z$:
\begin{equation}
    \lambda(x,y,z)=x^2+y^2+z^2-2(xy+xz+yz)\,.
\end{equation}
Note that the $\theta$-function ensures the positivity of $\lambda\left(p^2_1,m^2_1,m^2_2\right)$ and~\eqref{eq:D12term} is real as expected. 
In the case where either $m_1$ or $m_2$ are fakeons, the operation in \eqref{eq:B0f} simply removes the imaginary part of the $B_0$ function. Thus, we write
\begin{eqnarray}
B_0^{\rm f}(p^2_1,m^2_1,m^2_2)= 
\begin{cases}
 B_0(p^2_1,m^2_1,m^2_2) &\quad\text{if $m_1 \wedge m_2 \in $ standard particles}\\
 \text{Re}\left[B_0(p^2_1,m^2_1,m^2_2)\right] &\quad\text{if $m_1 \lor m_2 \in $ fake particles}. 
\end{cases}
\end{eqnarray}
Despite its simplicity, this modification gives nontrivial consequences. For example, it can change the width of an unstable particle, which might even become stable if it interacts with fakeons only. An explicit example is given in~\autoref{sec:toy}, where we show how $B_0^{\text{f}}$ modifies the peak of a standard particle.

The situation is more involved in the case of the $C_0$ and $D_0$ functions because of the presence of multiple thresholds, which in general introduce modifications in both real and imaginary parts of TV-functions.

\subsection{$C^{\rm f}_0$ function}\label{sec:C0f_function}

The one-loop scalar three-point function is written as $C_0(p^2_1,p^2_2,p^2_3,m^2_1,m^2_2,m^2_3)$, where $p_i$ are the momenta of the external legs and $m_i$ are the masses of the internal legs. Given the rotation symmetry of the triangle diagram, the $C_0$ function is symmetric under cyclic permutations of the indices of its arguments.
Therefore, to compute $\Delta_{\rm f}C_0$, it is sufficient to derive the terms
\begin{equation}
\label{eq:DQandDDterms}
    \int\frac{\mathrm{d}^3\boldsymbol{q}}{(2 \pi)^3}\frac{ \Delta^{ab}\mathcal{Q}^{ac}}{8\,\omega_a\omega_b\omega_c},\qquad \int\frac{\mathrm{d}^3\boldsymbol{q}}{(2 \pi)^3}\frac{ \Delta^{ab}\Delta^{ac}}{8\,\omega_a\omega_b\omega_c},
\end{equation}
for some of $a$,$b$ and $c$. We compute the case $(a,b,c)=(1,2,3)$ as detailed in~\autoref{appcomp}. Other $\Delta\mathcal{Q}$- and $\Delta\Delta$-terms appearing in \eqref{eq:DeltaC0} are those with $(2,3,1)$ and $(3,1,2)$. They can be easily deduced from the case $(1,2,3)$ by cyclically permuting the indices of all external momenta and masses. In this way we obtain three over six permutations that appear in~\eqref{eq:DeltaC0}. However, the remaining three permutations and those derived as explained above are mutually exclusive, since they are associated with cuts that have opposite shadings. Besides this detail, the expressions of the missing permutations are equal to those that we derive here.

In Lorentz invariant form, the $\Delta^{12} \mathcal{Q}^{13}$-term reads

\begin{equation}\label{eq:D12Q13LorInv}
\int\frac{\mathrm{d}^3\boldsymbol{q}}{(2 \pi)^3}\frac{ \Delta^{12}\mathcal{Q}^{13}}{8\,\omega_1\omega_2\omega_3} = \frac{\theta(p_1^2-(m_1+m_2)^2)}{16 \pi\sqrt{\lambda(p_1^2,p_2^2,p_3^2)}}\ln\left|\frac{u_{13}+1}{u_{13}-1}\right|, 
\end{equation}

\begin{equation}\label{eq:ctheta13}
    u_{13} =\frac{p_1^2(-p_1^2+p_2^2+p_3^2-2 m_3^2)+m_1^2(p_1^2+p_2^2-p_3^2)+m_2^2(p_1^2-p_2^2+p_3^2)}{\sqrt{\lambda(p_1^2,p_2^2,p_3^2)\lambda(p_1^2,m_1^2,m_2^2)}},
\end{equation}
while the $\Delta^{12}\Delta^{13}$-term is given by
\begin{eqnarray}\label{eq:D12D13LorInv}
 \int\frac{\mathrm{d}^3\boldsymbol{q}}{(2 \pi)^3}\frac{ \Delta^{12}\Delta^{13}}{8\,\omega_1\omega_2\omega_3} = \frac{\theta(p_1^2-(m_1+m_2)^2,p_3^2-p_2^2-m_1^2 + m_2^2,1-|u_{13}|)}{16 \sqrt{\lambda(p_1^2,p_2^2,p_3^2)}}.
\end{eqnarray}
The positivity of $\lambda (p_1,m^2_1,m_2^2)$ is ensured by the $\theta$-function, while that of $\lambda (p_1^2,p_2^2,p_3^2)$ is not. However, since we are considering only situations where $C_0^{\text{f}}$ is inserted in a one-loop amplitude, we have additional kinematical constraints on the Mandelstam variables, so that the case $\lambda(p_1^2,p_2^2,p_3^2)<0$ is excluded from the physical region. The domains where $\lambda(p_1^2,p_2^2,p_3^2)<0$ are considered in the study of analytic properties of the triangle diagram and its dispersion relations (see for example~\cite{Lucha:2006vc}). 
In the particular case of $\lambda(p_1^2,p_2^2,p_3^2)=0$, formula~\eqref{eq:D12D13LorInv} is divergent. This can be obtained, for example, in a $t$-channel diagram where $p^2_1=t$ and $p^2_2=p^2_3=m^2$. In that case the singular term is $\propto 1/\sqrt{t}$. Such behavior of the real part cannot occur in standard amplitudes, unless the particles in the loop are massless. In this respect, it is important to highlight that a singularity at $t=0$ proportional to $1/\sqrt{t}$ induces a long-range interaction effectively generated at one-loop order. This is of particular importance for model building using fakeons. In fact, models that enjoy such a feature could induce new-physics effects that are necessarily more constrained in order to reproduce the observed phenomenology.

The relation between \eqref{eq:D12Q13LorInv} and~\eqref{eq:D12D13LorInv} is evident, being both real functions. They can be written together as the real and imaginary parts of a more general function, which is proportional to one of the cut diagrams. This can be easily seen from the third column of~\autoref{ts1}\footnote{The cut diagrams in~\autoref{ts1} have two $\Delta\Delta$-terms, one of which is always zero, once the energy flow is fixed.}. This kind of relations between the terms of the decomposition are not manifest in the case of the box diagram (and higher-point functions) since certain identities~\cite{Anselmi:2021hab} between $\Delta\mathcal{P}\mathcal{P}$- and $\Delta\Delta\Delta$-terms are used to derive the threshold decomposition~\eqref{Boxdecomp}. 

The conditions in the $\theta$-function on the right-hand side of~\eqref{eq:D12D13LorInv} can be written as one of the following equivalent configurations, where the three external momenta of $C_0$ are constrained:
\begin{eqnarray}
\label{eq:condp2pm}
   && p_1^2 > (m_1+m_2)^2,\quad p_3^2 > (m_1+m_3)^2,\quad p^{2-}_{u 2} < p_2^2 < p^{2+}_{u 2},\\
\label{eq:condp3pm}
   && p^2_2 < (m_2-m_3)^2,\quad p^2_1 > (m_2+m_1)^2,\quad p^{2-}_{u 3} < p^2_3 < p^{2+}_{u 3},\\
\label{eq:condp1pm}
   &&  p^2_3 > (m_3+m_1)^2,\quad p^2_2 < (m_3-m_2)^2,\quad  p^{2-}_{u 1} < p^2_1 < p^{2+}_{u 1},
\end{eqnarray}
where
\begin{eqnarray}
\label{eq:pthetapm}
    p_{u 2}^{2\pm}=p_1^2+p_3^2-\frac{(p_1^2-m_2^2+m^2_1)(p_3^2-m_3^2+m^2_1)}{2 m_1^2}
    \pm\frac{\sqrt{\lambda(p_1^2,m_1^2,m_2^2)\lambda(p_3^2,m_1^2,m_3^2)}}{2 m_1^2},
\end{eqnarray}
and $p_{u 3}^{2\pm}$ and $p_{u 1}^{2\pm}$ are derived by means of the cyclic permutations (2,3,1) and (3,1,2) of the indices of the momenta and masses in $p_{u 2}^{2\pm}$. 
It is interesting to note that~\eqref{eq:condp2pm} implies $p_2^2<(m_2-m_3)^2$, while~\eqref{eq:condp3pm}, \eqref{eq:condp1pm} imply $p_1^2>(m_1+m_2)^2$ and $p_3^2>(m_1+m_3)^2$, respectively. Hence, a necessary condition for a $\Delta\Delta$-term to be nonvanishing is that two momenta be above their thresholds and the third one be below. The conditions~\eqref{eq:condp2pm} are more clear in this respect, since they show explicitly which external momenta have to be above threshold, while this information is not manifest in~\eqref{eq:condp3pm} and~\eqref{eq:condp1pm}. However, depending on the configuration, the conditions~\eqref{eq:condp3pm} and~\eqref{eq:condp1pm} can be used to better understand which terms are turned on (see below). 
\begin{figure}[t]
    \centering
    \includegraphics[scale=0.39]{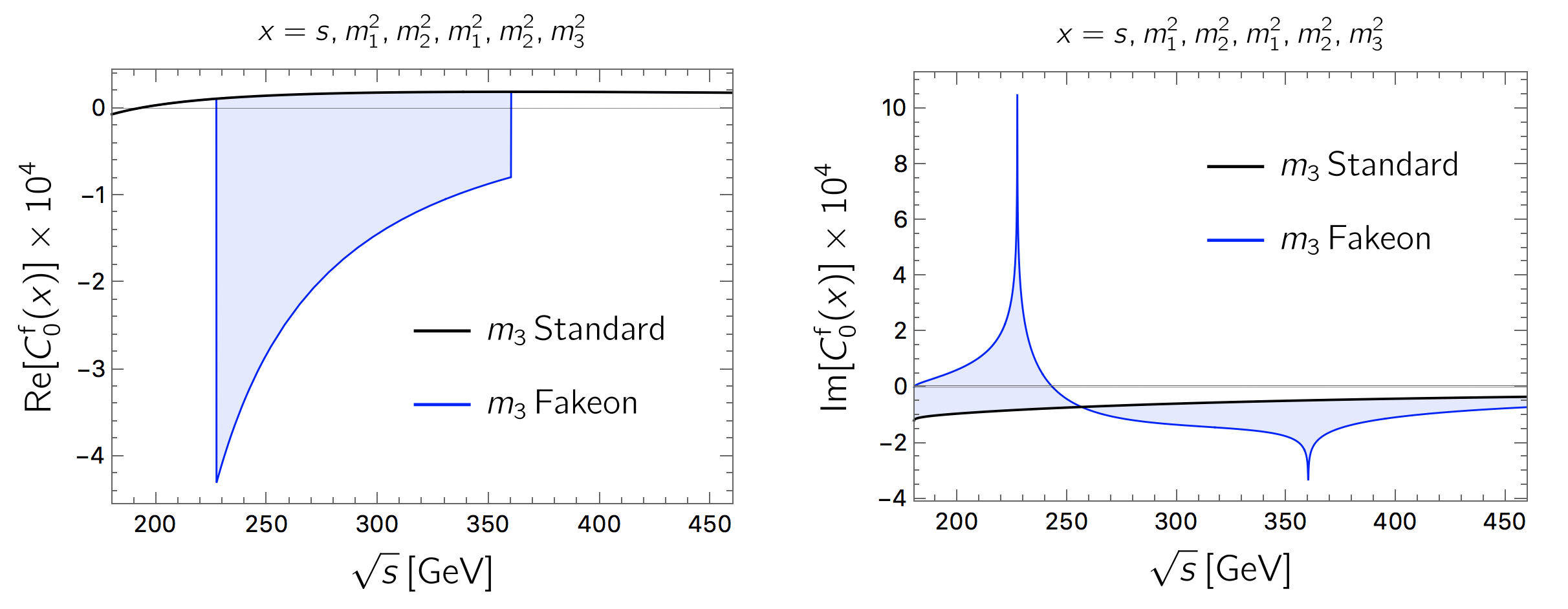}
    \caption{ Difference in the real and imaginary parts of the modified TV-function $C^{\rm f}_0$ in the case the $m_3$ is a standard particle (black line) or a fake particle (blue line). The values of the masses are $(m_1,m_2,m_3)=(5,170,80)$ GeV.}
    \label{fig:C0_example}
\end{figure}
The previous expressions simplifies in some particular configurations. For example, in the degenerate case $m_1=m_2=m_3=m$ we have
\begin{equation}
    p_1^2>4m^2, \quad p_3^2>4m^2,\quad p_{u2}^{2-}< p_2^2< p_{u2}^{2+},
\end{equation}    
    \begin{equation}
    p_{u2}^{2\pm}=p_1^2+p_3^2-\frac{p_1^2p_3^2}{2 m^2}\pm\frac{\sqrt{p_1^2(p_1^2-4m^2)}\sqrt{p_3^2(p_3^2-4m^2)}}{2 m^2},
\end{equation}
in agreement with~\cite{Lucha:2006vc}. In the massless case, it further reduces to
\begin{equation}
    p_1^2>0,\quad p_3^2>0, \quad p_2^2< 0. 
\end{equation}
The last configuration can be obtained when a triangle diagram is part of a scattering process in the $t$- or $u$-channel.

In Figure \ref{fig:C0_example} we illustrate how the real and imaginary parts of a $C_0$ function are modified in the presence of a fakeon for a specific example. We consider three particles of masses $m_1,m_2,m_3$, where $m_3$ can be either a standard particle or a fakeon and compare the results. In order to obtain a configuration in which both the $\Delta\mathcal{Q}$ and $\Delta\Delta$-terms are nonvanishing we impose $m_2>m_1+m_3$. For illustrative purposes, we choose the values $(m_1,m_2,m_3)=(10,170,80)$ GeV. Moreover, we set two external momenta on shell, i.e. $p_2^2=m_1^2$ and $p_3^2=m_2^2$, while $p_1^2=s$ is the center-of-mass energy squared. These choices mimic a physical situation where the $C_0$ function is inserted in a scattering amplitude. Then, the conditions~\eqref{eq:condp1pm} turns into
\begin{equation}\label{examplecond}
    m_2^2>(m_3+m_1)^2, \qquad m_1^2<(m_3-m_2)^2, \qquad 2(m_1^2+m_2^2)-m_3^2<s<\frac{(m_2^2-m_1^2)^2}{m_3^2}.
\end{equation}
The first and the second conditions are always satisfied, while the third one depends on $s$. From the left panel in~\autoref{fig:C0_example} we can see that the $\Delta\Delta$-term is turned on when $\sqrt{s}$ reaches $\sqrt{p_{u 1}^{2-}}\simeq 227$ GeV and disappear when $\sqrt{s}$ reaches $\sqrt{p_{u 1}^{2+}}\simeq 360$ GeV. Moreover, since the first condition in~\eqref{examplecond} involves $m_3^2$, there is a difference also in the imaginary part, as depicted in the right panel in~\autoref{fig:C0_example}. Both the real and the imaginary parts have nonanalyticities in correspondence of $p_{u 1}^{2\pm}$.

\begin{figure}[t]
    \centering
    \includegraphics[scale=0.39]{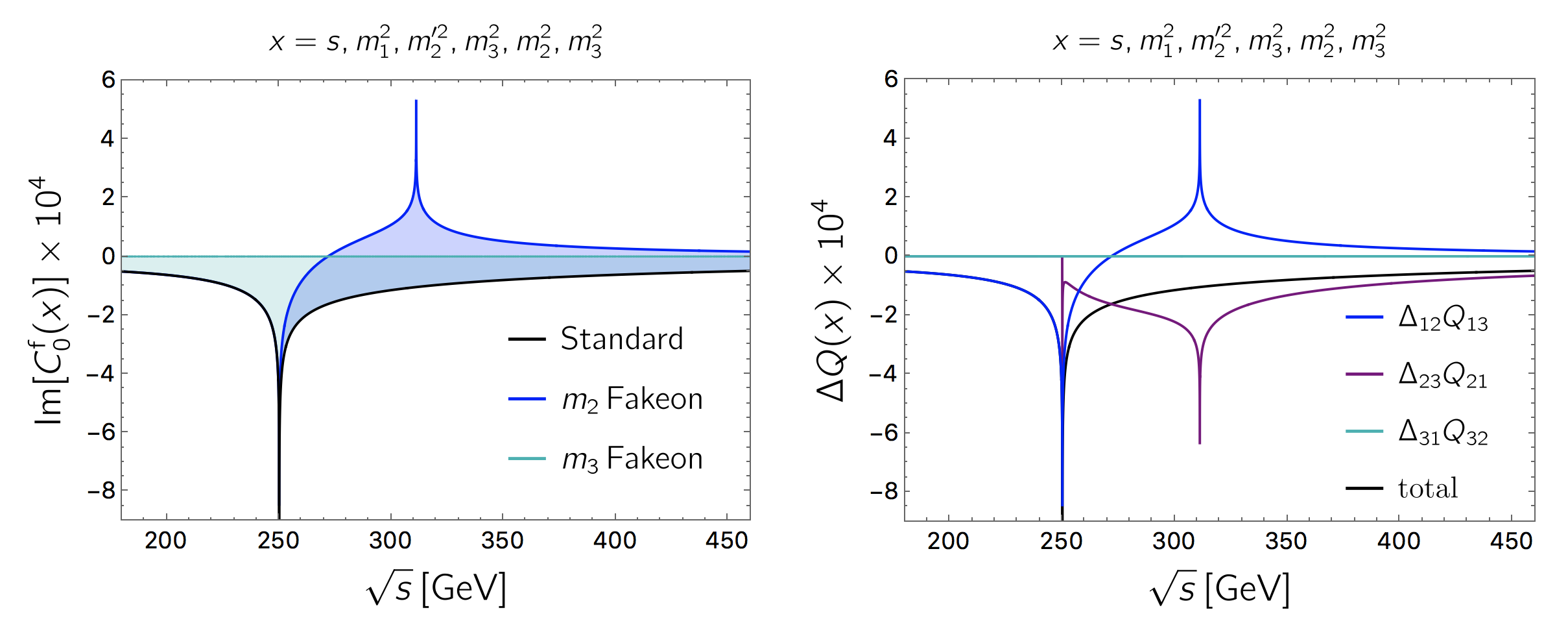}
    \caption{ Left panel: difference in the imaginary part of the modified TV-function $C^{\rm f}_0$ (filled regions) when different particles in the loop are turned into fakeons. The exemplifying values chosen for the masses are $(m_1,m_2=m'_2,m_3)=(5,170,80)$ GeV. Right panel: separate $\Delta \mathcal{Q}$-terms that contribute to the total imaginary part of a standard $C_0$ function. }
    \label{fig:C0TS_example}
\end{figure}

To summarize this example, in the case of standard particles the triangle diagram has two external momenta above threshold, hence a nonzero imaginary part. In the case where $m_3$ is a fakeon one of the two contributions to the imaginary part vanishes and for every $s>(m_1+m_2)^2$ there is a difference. Moreover, when $s$ is between the critical values $p_{u 1}^{2\pm}$ there is also a difference in the real part.

The new singularities cannot be physically interpreted as it is usually done in the standard cases because of their different nature. Typically, a physical interpretation is provided by the Coleman-Norton theorem~\cite{Coleman:1965xm}, which associates singularities of Feynman amplitudes with processes described by the diagrams where all the particles are on shell. This description cannot be applied when fakeons circulate in the loop, since they cannot be on shell by construction and, therefore, the processes that the theorem would associate with the singularities do not exist. The reason is that by choosing a different prescription we violate one hypothesis of the Coleman-Norton theorem, i.e. the assumption that the Feynman prescription is adopted for each propagator.

Features such as those in~\autoref{fig:C0_example} are due to the fact that the fakeon prescription sets some cut diagrams to zero. A single cut diagram can have singularities that cancel each other in the sum of all the cut diagrams. If some of them are set to zero while the others do not, additional singularities might appear. In order to highlight this in~\autoref{fig:C0TS_example} we give an example of a triangle diagram that has a singularity in the imaginary part already in the standard case (black line), a so-called \emph{triangle singularity}~\cite{Landau:1959fi}, see also e.g.~\cite{Du:2019idk}. Interestingly, the singularity is entirely removed when at least two of the particles in the internal lines are turned into fakeons (light-green line). In this case none of the on-shell processes required by the Coleman-Norton theorem can take place. More involved is the case where only particle $m_2$ is a fakeon (blue line), in which a new singularity appear. Its mathematical origin is understood from the right panel of~\autoref{fig:C0TS_example} where we plot separately the real parts of the three cut diagrams. We see that two of them (blue and purple lines) have an opposite singularity that cancels in the sum. Therefore, it does not appear in the standard case, while it does when only one of the two $\Delta \mathcal{Q}$-term is removed by the fakeon prescription.
The new singularities deserve a physical interpretation by means of a suitable generalization of the Coleman-Norton theorem, where both standard particles and fakeons are taken into account.

\subsection{$D^{\rm f}_0$ function}\label{sec:D0f_function}
The scalar four-point function is written as $D_0(p^2_1,p^2_2,p^2_3,p^2_4,p^2_{12},p^2_{23},m^2_1,m^2_2,m^2_3,m^2_4)$, where $p_i$ are the momenta of the external legs, $p_{ij}^2\equiv (p_i+p_j)^2$, and $m_i$ are the masses of the internal legs. The $D_0$ function is symmetric under cyclic permutations of the indices of its arguments. Moreover, it is invariant under reflections along two opposite internal lines, such as $(p^2_1\leftrightarrow p^2_4 , p^2_2\leftrightarrow p^2_3, m^2_2\leftrightarrow m^2_4)$, or twists of two adjacent internal lines, for example $(p^2_1\leftrightarrow p^2_{12},p^2_3\leftrightarrow p^2_{23}, m^2_2\leftrightarrow m^2_3)$, and every cyclic permutation of them. 
Making use of these symmetry properties, to compute $\Delta_{\rm f}D_0$ it is sufficient to derive the terms
\begin{equation}
    \int\frac{\mathrm{d}^3\boldsymbol{q}}{(2 \pi)^3}\frac{ \Delta^{ab}\mathcal{Q}^{ac}\mathcal{Q}^{ad}}{16\,\omega_1\omega_2\omega_3\omega_4},\quad \int\frac{\mathrm{d}^3\boldsymbol{q}}{(2 \pi)^3}\frac{ \Delta^{ab}\Delta^{ac}\mathcal{Q}^{ad}}{16\,\omega_1\omega_2\omega_3\omega_4},\quad \int\frac{\mathrm{d}^3\boldsymbol{q}}{(2 \pi)^3}\frac{ \Delta^{ab}\Delta^{ac}\Delta^{ad}}{16\,\omega_1\omega_2\omega_3\omega_4}\,,
\end{equation}
for some $a$,$b$, $c$ and $d$. We choose the case $(a,b,c,d)=(1,2,3,4)$ as detailed in \autoref{appcomp}. In $D_0$ there are 6 nonequivalent terms of the $\Delta\mathcal{Q}\mathcal{Q}$-type. All of them can be easily deduced from the case $(1,2,3,4)$. In particular,  $(2,3,4,1)$, $(3,4,1,2)$ and $(4,1,2,3)$ are obtained by cyclically permuting the indices of all the momenta and masses appearing in~\eqref{eq:DQQBox}-\eqref{eq:calpha24Box}. The cases $(1,3,2,4)$ and $(4,2,3,1)$ are obtained by means of twists of the internal lines 2, 3 and 1, 4, respectively. As in the case of the triangle diagram, there are more permutations in~\eqref{eq:DeltaD0} but half of them are mutually exclusive due to opposite shadings in the cut diagrams and the permutations mentioned above are sufficient to cover all the cases. Analogous considerations apply also for the $\Delta\Delta\mathcal{Q}$- and $\Delta\Delta\Delta$-terms.

The $\Delta^{12} \mathcal{Q}^{13}\mathcal{Q}^{14}$-term reads
\begin{eqnarray}
\label{eq:DQQBox}
   \int\frac{\mathrm{d}^3\boldsymbol{q}}{(2 \pi)^3}\frac{ \Delta^{12}\mathcal{Q}^{13}\mathcal{Q}^{14}}{16\,\omega_1\omega_2\omega_3\omega_4} 
   = \frac{p^2_{1}\,\theta\left(\,p_{1}^2-(m_1+m_2)^2\,\right)}{8 \pi\sqrt{\lambda(p^2_1,p^2_2,p^2_{12})\lambda(p^2_1,p^2_4,p^2_{23})\lambda(p^2_{1},m^2_1,m^2_2)\left|\kappa(u_{13},u_{14},c_{24})\right|}}\hspace{0.5cm}
\end{eqnarray}
\newcommand{\sdfrac}[2]{\mbox{\small$\displaystyle\frac{#1}{#2}$}}
\begin{eqnarray}
\times
\begin{cases}
 \text{sgn}(\beta^+) \, \text{ln}\left|\sdfrac{\kappa-\alpha(u_{13}-1)+|\beta^+|\sqrt{\kappa} }{\delta(u_{13}-1)} \right| -\text{sgn}(\beta^-)\,\text{ln}\left|\sdfrac{\kappa-\alpha(u_{13}+1) +|\beta^-| \sqrt{\kappa}}{\delta(u_{13}+1)}\right|  & \text{if $\kappa>0$ }\nonumber\\[15pt]
\text{sgn}(\beta^-)\,\text{arcsin}\left[\sdfrac{\kappa-\alpha(u_{13}+1)}{ \delta|u_{13}+1|}\right]  - \text{sgn}(\beta^+)\,\text{arcsin}\left[\sdfrac{\kappa-\alpha(u_{13}-1) }{ \delta|u_{13}-1|}\right] & \text{if $\kappa< 0$} \\[10pt]
+\, \text{sgn}(\beta^-)\,\theta\left(|c_{24}|-|u_{14}|\right)\pi   
\end{cases}\nonumber \\[-15pt]
\end{eqnarray}
where we define the function 
\begin{equation}
    \kappa(x,y,z)=x^2+y^2+z^2+2 xyz-1,
\end{equation}
and we introduce the quantities
\begin{eqnarray}
\label{eq:alphabetadelta}
  & \alpha  =u_{13} + u_{14} c_{24},\quad \beta^{\pm}=u_{14} \pm c_{24} , \\[5pt]
  & \delta = \sqrt{(1-u^2_{14})(1-c^2_{24})}\, , \quad \kappa=\kappa(u_{13},u_{14},c_{24})=\alpha^2-\delta^2.
\end{eqnarray}
The term $u_{13}$ coincides with~\eqref{eq:ctheta13} once we replace $p^3_3 \rightarrow p^2_{12}$, while $u_{14}$ and $c_{24}$ are
\begin{eqnarray}
    \label{eq:ctheta14Box}
    u_{14} &=&\frac{p_1^2(-p_1^2+p_{23}^2+p_{4}^2-2 m_4^2)+m_1^2(p_1^2+p_{23}^2-p_{4}^2)+m_2^2(p_1^2-p_{23}^2+p_{4}^2)}{\sqrt{\lambda(p_1^2,p_{4}^2,p_{23}^2)\lambda(p_1^2,m_1^2,m_2^2)}}\,,\\
     \label{eq:calpha24Box}
    c_{24} &=&\frac{p_1^2(-p_1^2+p_{23}^2+p_4^2-2 p_3^2)+p_{12}^2(p_1^2+p_{23}^2-p_4^2)+p_2^2(p_1^2-p_{23}^2+p_4^2)}{\sqrt{\lambda(p_1^2,p_4^2,p_{23}^2)\lambda(p_1^2,p_2^2,p_{12}^2)}} \,,
\end{eqnarray}
which again can be obtain from~\eqref{eq:ctheta13} with suitable substitutions. Note that $c_{24}$ is the cosine of the kinematic angle between $\boldsymbol{p}_2$ and $\boldsymbol{p}_4$ in the frame where $\boldsymbol{p}_1=0$. Thus its expression involves the external momenta only and we always have $|c_{24}|< 1$ for physical values of $p_i$. 

The terms $\Delta^{12}\Delta^{13}\mathcal{Q}^{14}$ and $\Delta^{12}\Delta^{13}\Delta^{14}$ read
\begin{eqnarray}
\label{eq:DDQBox}
    \int\frac{\mathrm{d}^3\boldsymbol{q}}{(2 \pi)^3}\frac{ \Delta^{12}\Delta^{13}\mathcal{Q}^{14}}{16\,\omega_1\omega_2\omega_3\omega_4} &=& \frac{p^2_{1}\,\theta\left(\,p_{1}^2-(m_1+m_2)^2,p_{12}^2-p_2^2-m_1^2 + m_2^2,1-|u_{13}|\,\right)}{8\,\sqrt{\lambda(p^2_1,p^2_2,p^2_{12})\,\lambda(p^2_1,p^2_4,p^2_{23})\,\lambda(p^2_{1},m^2_1,m^2_2)}}\,\nonumber\\[5pt]
    &\times& \frac{\text{sgn}(u_{14})\theta\left(\,\kappa(u_{13},u_{14},c_{24})\, \right) }{\sqrt{\kappa(u_{13},u_{14},c_{24})}}.
    \\[5pt]
\label{eq:DDDBox1}
    \int\frac{\mathrm{d}^3\boldsymbol{q}}{(2 \pi)^3}\frac{ \Delta^{12}\Delta^{13}\Delta^{14}}{8\,\omega_1\omega_2\omega_3\omega_4}
    &=& \frac{p^2_{1}\,\theta\left(\,p_{1}^2-(m_1+m_2)^2,p_{12}^2-p_2^2-m_1^2 + m_2^2,1-|u_{13}|\,\right)}{8\,\sqrt{\lambda(p^2_1,p^2_2,p^2_{12})\,\lambda(p^2_1,p^2_4,p^2_{23})\,\lambda(p^2_{1},m^2_1,m^2_2)}}\nonumber\\[5pt]
\label{eq:DDDBox2}
    &\times& \frac{\theta\left(\,p^2_4-p^2_{23}-m^2_2+m^2_1, -\kappa(u_{13},u_{14},c_{24}) \right) }{\sqrt{-\kappa(u_{13},u_{14},c_{24})}}.
\end{eqnarray}
These two terms cannot appear together in the same diagram, since they require disjoint conditions to be both nonvanishing. In particular, the sign of the function $\kappa$ is opposite in the two cases. 
\begin{figure}[t]
    \centering
    \includegraphics[scale=0.4]{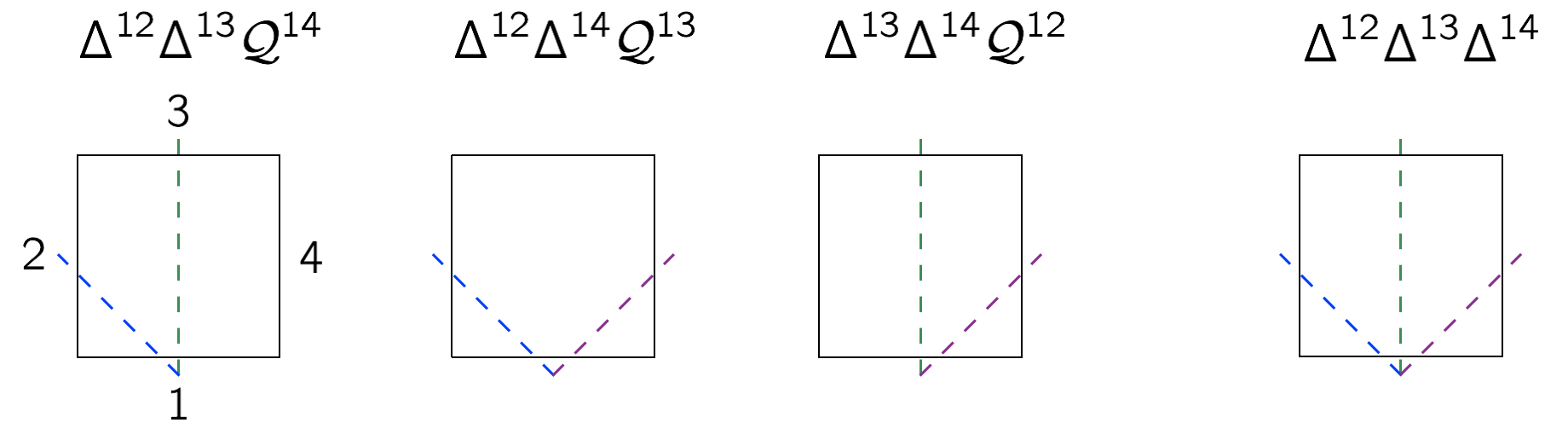}
    \vspace{-2mm}
    \caption{Basic types of $\Delta\Delta\mathcal{Q}$- and $\Delta\Delta\Delta$- terms appearing in $D_0$, the rest are obtained by cyclically permuting all the indices of momenta and masses. The dotted lines indicate the thresholds that must be exceeded by the corresponding invariants built with external momenta.}
    \label{fig:Main_DDQandDDD_Box}
\end{figure}
The conditions imposed by the $\theta$-functions in~\eqref{eq:DDQBox} and~\eqref{eq:DDDBox1} can be reformulated in the same fashion of~\eqref{eq:condp2pm} as follows. First, note that one of the two $\theta$-functions, which appears in both terms, has the same form of that in~\eqref{eq:D12D13LorInv}. Therefore, it gives a set of conditions similar to those in~\eqref{eq:condp2pm}. Then, since $|u_{13}|< 1$ and $|c_{24}|< 1$, it is easy to check that $\kappa\lessgtr 0$ is equivalent to
\begin{eqnarray}
\label{eq:Kbounds}
|u_{14} + u_{13}c_{24}| \lessgtr \sqrt{(1-u^2_{13})(1-c^2_{24})}\, .
\end{eqnarray}
Moreover, requiring $\kappa <0$ implies $|u_{14}|< 1$, which, together with $p^2_4-p^2_{23}-m^2_2+m^2_1> 0$, gives a condition analogous to the third one in~\eqref{eq:condp2pm} for $p_{23}^2$ and that $p_4^2$ be above threshold. Finally, \eqref{eq:Kbounds} can be rewritten as a constraint on $p^2_3$. To summarize, the configurations that provide a nonvanishing $\Delta^{12}\Delta^{13}\mathcal{Q}^{14}$ are
\begin{eqnarray}\label{eq:DDQcond}
   & p^2_1 > (m_1+m_2)^2 , \quad p^2_{12} > (m_1+m_3)^2 , \quad p^{2-}_{u 2} < p^2_{2} < p^{2+}_{u 2} , \quad p^2_{3}<p^{2-}_{v 3} \lor p^2_{3}>p^{2+}_{v 3}  \,,
\end{eqnarray}
while those for a nonvanishing $\Delta^{12}\Delta^{13}\Delta^{14}$ are
\begin{eqnarray}\label{eq:DDDcond}
   & p^2_1 > (m_1+m_2)^2 , \quad p^2_{12} > (m_1+m_3)^2 , \quad p^2_4 > (m_1+m_4)^2 , \nonumber\\
   & p^{2-}_{u 2}< p^2_{2} < p^{2+}_{u 2}, \quad p^{2-}_{u 23} < p^2_{23} < p^{2+}_{u 23} , \quad p^{2-}_{v 3} < p^2_{3} < p^{2+}_{v 3}  \,,
\end{eqnarray}
 where $p^{2\pm}_{u 2}$ and $p^{2\pm}_{u 23}$ are obtained from~\eqref{eq:pthetapm} applying the substitutions $(p^2_3\rightarrow p^2_{12})$ and $(p^2_3\rightarrow p^2_{4},m^2_3\rightarrow m^2_{4})$, respectively, while  $p^{\pm}_{v 3}$ is given by
\begin{eqnarray}
\label{eq:Mphipm}
p_{v 3}^{2\pm}&=&p_{12}^2+p_{4}^2-\frac{(p_{12}^2-p_2^2+p^2_1)(p_4^2-p_{23}^2+p^2_1)}{2 p_{1}^2}\nonumber\\
    &+&\left(u_{13} u_{14}\pm \sqrt{(1-u^2_{13})(1-u^2_{14})}\right)\frac{\sqrt{\lambda(p_1^2,p_2^2,p_{12}^2)\lambda(p_1^2,p_4^2,p_{23}^2)}}{2 p_{1}^2}.
\end{eqnarray}
Again, the conditions~\eqref{eq:DDQcond} and~\eqref{eq:DDDcond} can be rewritten in other equivalent ways. We choose to show those where it is possible to explicitly read which momenta must be above threshold. Note that, as in the case of $C_0^{\text{f}}$, the term with the maximum number of $\Delta$'s constrains all the six invariants in the argument of $D_0^{\text{f}}$.

\begin{figure}[t]
    \centering
    \includegraphics[scale=0.39]{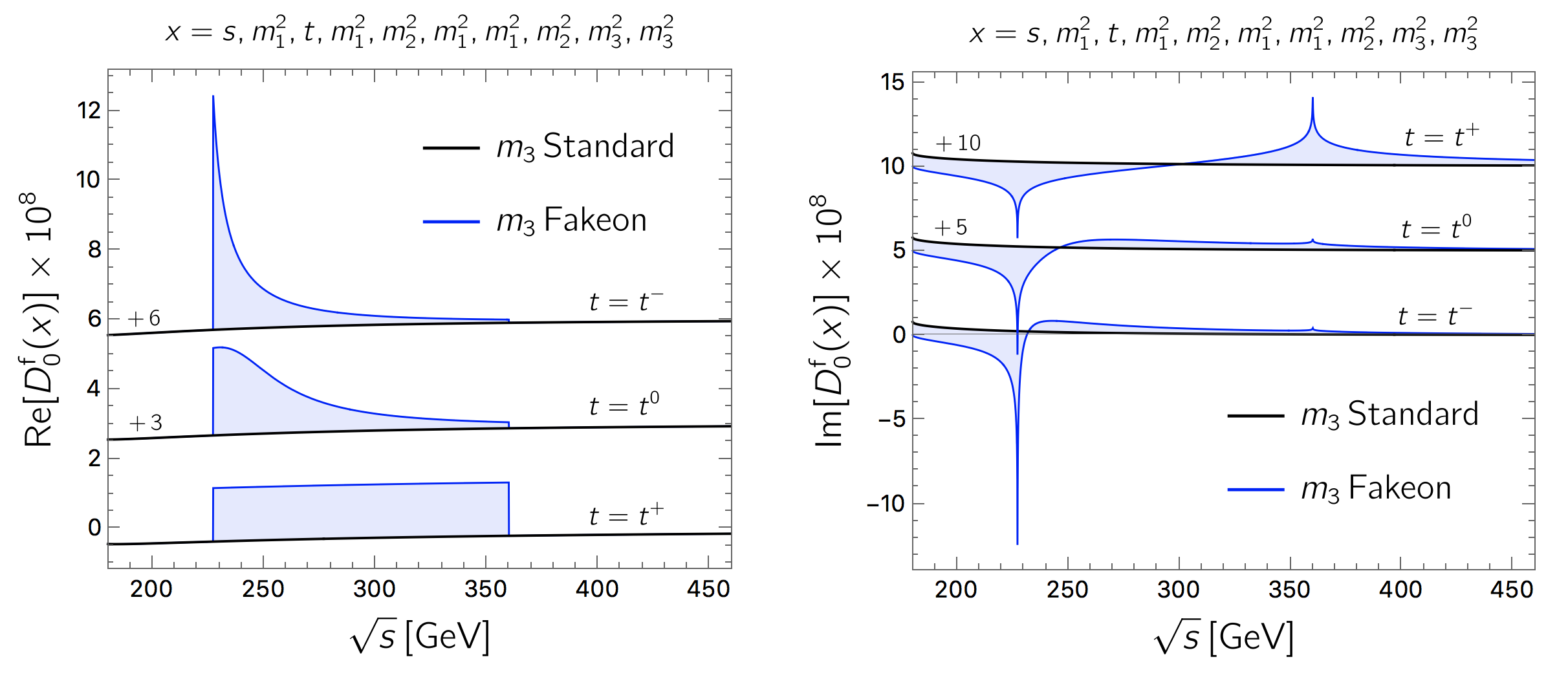}
    \captionsetup{format=hang}
    \caption{Difference in the real and imaginary parts of $D^{\rm f}_0$ as a function of the center-of-mass energy $\sqrt{s}$ in the case where $m_3$ is a standard particle (black line) or a fake particle (blue line). We choose the values $(m_1,m_2,m_3)=(10,170,80)$ GeV. The difference is displayed for three representative values of $t=(t^-,t^0,t^+)$ to highlight the dependence on $t$. An offset is applied to visually separate the three cases.}
    \label{fig:D0_example}
\end{figure}

\autoref{fig:Main_DDQandDDD_Box} shows the types of $\Delta\Delta\mathcal{Q}$- and $\Delta\Delta\Delta$-terms contained in a $D_0$ function. The full set is the collection of all the cyclical permutations of the external momenta and masses of those basic types. The dashed lines indicate which threshold must be exceeded by the corresponding invariants built with external momenta. Since the $\Delta$'s appearing in these terms share one common index, the invariants that must be above threshold are always associated with cuts that intersect one common line. Graphically, this results in a ``V-like shape" threshold configuration in the case of $\Delta\Delta\mathcal{Q}$-terms, and in a ``arrow-like shape" configuration in the case of $\Delta\Delta\Delta$-terms. Other configurations are vanishing\footnote{We stress that the configurations depicted in~\autoref{fig:Main_DDQandDDD_Box} are necessary but not sufficient in order to have nonvanishing $\Delta\Delta\mathcal{Q}$- and $\Delta\Delta\Delta$-terms.}.
Finally, we give a concrete example. In~\autoref{fig:D0_example} we show the real and imaginary parts of a $D_0^{\text{f}}$. In particular, we consider the same masses used in~\autoref{fig:C0_example}, i.e. $(m_1,m_2,m_3)=(10,170,80)$ GeV, and set, $p_2^2=p_4^2=p_{23}^2=m_1^2$ and $p_{12}^2=m_2^2$, while $p_{1}^2=s$ and $p_{3}^2=t$ are the Mandelstam variables.
In this setting, the $\Delta\Delta\Delta$-term is absent, while both the $\Delta\mathcal{Q}\mathcal{Q}$ and $\Delta\Delta\mathcal{Q}$ are nonvanishing. Hence, the modifications occur both in the real and imaginary parts. 
The first three conditions in~\eqref{eq:DDQcond} are the same as in the case of the triangle diagram. Therefore, the difference is present in the same range $227 \ \text{GeV}\lesssim \sqrt{s} \lesssim 360 \ \text{GeV}$. The fourth condition in~\eqref{eq:DDQcond} bounds the values of $t$ as $s$ varies. To highlight the dependence on $t$, in this example we fix $t$ to three representative ($s$-dependent) values $t^+,t^0,t^-$, obtained from the formula that relates $t$ with the scattering angle $\theta$ in the center-of-mass frame
\begin{eqnarray}
     t=\frac{(3m_1^2+m_2^2-s)}{2}+\cos\theta\frac{\sqrt{\lambda(s,m_1^2,m_1^2)\lambda(s,m_1^2,m_2^2)}}{2 s}
\end{eqnarray}
choosing $\cos\theta=1,0,-1$, respectively. Finally, the $\Delta\mathcal{Q}\mathcal{Q}$-term is present for $s>(m_1+m_2)^2$, which is always satisfied in the range $180 \ \text{GeV}\leq \sqrt{s}\leq 450 \ \text{GeV}$ that we chose in this example. Therefore, the imaginary part is always nonvanishing in this range and the fakeon case is always different from the standard one, since the only nonzero cut diagram involves $m_3$ in the threshold. As in the standard TV-functions, the nonanalyticities in the real part are located at the same point of those in the imaginary part. 

In~\autoref{sec:toy} we give an illustrative application by using a toy model, where the modifications described above appear in the total cross sections.

\subsection{Reducible functions}
The reducible functions are given by Feynman integrals with nontrivial numerators, i.e. any tensor monomial built with momenta and the metric. The reducible functions can be decomposed in terms of the TV-functions. For example, the integral defined by
\begin{equation}\label{eq:B1}
    B^{\mu}(p_1^2,m_1,m_2)\equiv\int\frac{\mathrm{d}^Dq}{(2 \pi)^D}\frac{q^{\mu}}{(q^2-m_1^2+i\epsilon)[(p+q)^2-m_2^2+i\epsilon]}
\end{equation}
can be written, by Lorentz invariance, as 
\begin{equation}
    B^{\mu}(p_1^2,m_1,m_2)=p^{\mu}B_1(p_1^2,m_1,m_2).
\end{equation}
Then, the scalar function $B_1$ is derived in terms of the TV-functions by contracting the integral with $p_{\mu}$. The result reads 
\begin{equation}\label{eq:B1red}
    B_1(p_1^2,m_1,m_2)=-\frac{1}{2p^2}\left[\left(p_1^2+m_1^2-m_2^2\right)B_0(p_1^2,m_1,m_2)-A_0(m_1)+A_0(m_2)\right].
\end{equation}
This operation is called Passarino-Veltman reduction~\cite{Passarino:1978jh}. Note that~\eqref{eq:B1red} is regular in $p^2=0$ and has the same branch cut structure of $B_0$. This is due to the fact that the numerator in the integral~\eqref{eq:B1} does not modify the location of the poles of the integrand. This holds for every Feynman integral with nontrivial numerators. 

Labelling the scalar reducible functions as $B_r$, $C_r$, etc., where $r$ is a number that uniquely identify the numerator of the integrals, their Passarino-Veltman reduction can be written as
\begin{equation}\label{eq:PVredB}
    B_r=g^{B}_{r}B_0+\sum_{i=1}^2f_{r,i}^{B}A_0(m_i)
\end{equation}
\begin{equation}\label{eq:PVredC}
    C_r=h_r^CC_0+\sum_{i,j,k=1}^3g_{r,ijk}^CB_0(p_i^2,m_j,m_k)+\sum_{i=1}^3f_{r,i}^CA_0(m_i),
\end{equation}
and so on, where the functions $f,g,h\ldots$ depend on momenta and masses and do not introduce any new singularities. Therefore, the reducible functions inherit the analytic structure from the TV ones. We conclude that the Passarino-Veltman reduction in the presence of fakeons is the same as in the standard case with the replacements $B_0\rightarrow B_0^{\text{f}}$, $C_0\rightarrow C_0^{\text{f}}$ etc.

\sect{Cross-section modifications}\label{sec:toy}
In the previous section we have shown that the presence of fakeons can affect both the the real and imaginary parts of the one-loop integrals. Nevertheless, the one-loop cross section is given by
 \begin{equation}
 \sigma=\int\mathrm{d}\Pi\Big[|\mathcal{A}_{\text{tree}}|^2+2\,\mathrm{Re}\left(\mathcal{A}_{\text{tree}}^*\mathcal{A}_{1\text{-loop}}\right)\Big],
 \end{equation}
 where $\mathrm{d}\Pi$ is the measure of the phase space of outgoing particles and $\mathcal{A}_{\text{tree},1\text{-loop}}$ are the tree-level and one-loop amplitudes, respectively. Since $\mathcal{A}_{\text{tree}}$ is real everywhere but at the poles of the propagators\footnote{Strictly speaking, $\mathcal{A_{\text{tree}}}$ is a complex distribution, because of the presence of the Feynman prescription. However, the $i\epsilon$ is unnecessary everywhere but at the poles.}, the second term in the squared bracket can be written as $2 \mathcal{A}_{\text{tree}}\mathrm{Re}(\mathcal{A}_{1\text{-loop}})$ almost everywhere. Therefore, modifications of the imaginary parts in $\mathcal{A}_{1\text{-loop}}$ do not affect the cross section at one loop. Thus, the differences between fakeons and standard particles can only occur through the $\Delta\Delta$- and $\Delta\Delta\mathcal{Q}$-terms coming from the triangle and box diagrams, respectively, provided that the configurations \eqref{eq:condp2pm} and \eqref{eq:DDQcond} are realized.
 
 A special remark should be made in the case of a scattering process mediated by an unstable particle with mass $M$. We can restrict the domain of the center-of-mass energy to a neighborhood of $M$, where the contribution of the triangle and box diagrams is negligible, and resum the self energies into the dressed propagator. Plugging it into the tree-level diagrams, $\mathcal{A}_{\text{tree}}$ acquires an imaginary part so the cross section becomes sensible to the modifications of the imaginary parts of $B_0^{\text{f}}$. 
 To quantify the difference between the standard case and the fakeon one we introduce the ratio
 \begin{eqnarray}
 \label{eq:dsigmaf}
\delta_{\rm f} \sigma\equiv \frac{\sigma_{\rm f}-\sigma}{\sigma},
 \end{eqnarray}
 where $\sigma_{\rm f}$ denotes the one-loop total cross section in the fakeon case where all the TV-functions are replaced by those in \eqref{eq:A0f}-\eqref{eq:D0f}. We emphasize that, since both $\sigma$ and $\sigma_{\text{f}}$ are linear in the one-loop amplitudes, in the regions away from the poles of the propagators in $\mathcal{A}_{\rm tree}$, only the sum of all ${\rm Re}(\Delta_{\rm f}C_0)$ and ${\rm Re}(\Delta_{\rm f}D_0)$ contributes in the ratio $\delta_{\text{f}}\sigma$. On the contrary, in the proximity of the poles of $\mathcal{A}_{\rm tree}$ the main contribution to $\delta_{\text{f}}\sigma$ comes from $\Delta_{\rm f}B_0$ in the dressed propagator.  
 In what follows we show two examples where these situations can be realized by means of a toy model equipped with the necessary features to illustrate the effects of the modified the TV-functions introduced in the previous section. In particular, we need at least three fields with a mass hierarchy such that, given the form of the interactions, all the possible decays of the heaviest one are allowed. Then one of the lighter fields is a fakeon, while the other two are standard particles. Moreover, we include a trilinear interaction that involves all three fields in order to easily have contributions coming from $\Delta\Delta$- and $\Delta\Delta\mathcal{Q}$-terms. 
 
 The field content of the model is given by three real scalar fields $\phi_i$ ($i=1,2,3$). We impose a discrete $\mathbb{Z}_2$ symmetry under which $\phi_1 \rightarrow -\phi_1$ and $\phi_2 \rightarrow -\phi_2$, while $\phi_3$ is even. This limits the number of trilinear couplings. Moreover, we are free to set the quartic couplings to zero, since they are not generated by renormalization. The most general Lagrangian reads
\begin{eqnarray}
\mathcal{L}&=&\sum^3_{i=1} \left[\frac{1}{2}(\partial_\mu \phi_i)^2 - \frac{m^2_i}{2} \phi_i^2\right]- \frac{\lambda_{3}}{3!}\phi^3_3 - \frac{\lambda_{13}}{2}\,\phi_1^2\phi_3- \frac{ \lambda_{23}}{2}\,\phi_2^2\phi_3-\lambda_{123}\,\phi_1\phi_2\phi_3.
\end{eqnarray}
   \begin{figure}[t]
    \centering
    \includegraphics[scale=0.28]{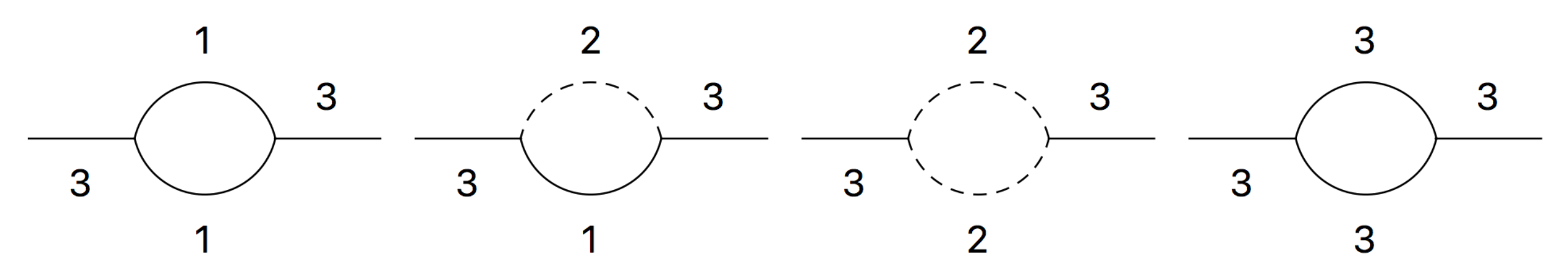}
    \captionsetup{format=hang}
    \caption{Self-energy diagrams of $\phi_3$. The dashed line identifies the fakeon. In the standard scenario the first three diagrams contribute to the decay width of $\phi_3$, while in the case where $\phi_2$ is a fakeon only the first one gives a non vanishing contribution to the width.}
    \label{fig:Gamma3_SelfEnergies}
\end{figure}  
In this framework we compute the scattering cross sections of different processes up to one loop. Renormalization is performed in the on-shell scheme where the mass and field renormalization constants, $\delta m^2_i$ and $\delta Z_i$, are computed from the unrenormalized self energies $\Sigma_i$ imposing the conditions $\delta m^2_i= \text{Re}[\Sigma_i (m^2_i)]$ and $\delta Z_i= -\text{Re}[\partial\Sigma_i (p^2)/\partial p^2]_{p^2=m^2_i}$.

 \subsection{Modified peak for standard particles}
 First we show how the presence of a fakeon can affect the peak of a standard, unstable particle.  Considering $\phi_2$ a fakeon, in order to obtain such modification we assume a mass hierarchy $m_3>m_1+m_2$ so all the possible decays of $\phi_3$ are allowed. In particular, we set $(m_1, m_2, m_3)=(10, 80, 170)$ GeV. Then we compute the cross section for the process $\phi_1\phi_1\rightarrow\phi_1\phi_1$ as a function of $\sqrt{s}$. Being $\phi_1$ the lightest particle, the configurations in \eqref{eq:condp1pm} and \eqref{eq:DDDcond} are never realized because they require two momenta above threshold and one below. 
 \begin{figure}[t]
    \centering
    \includegraphics[scale=0.325]{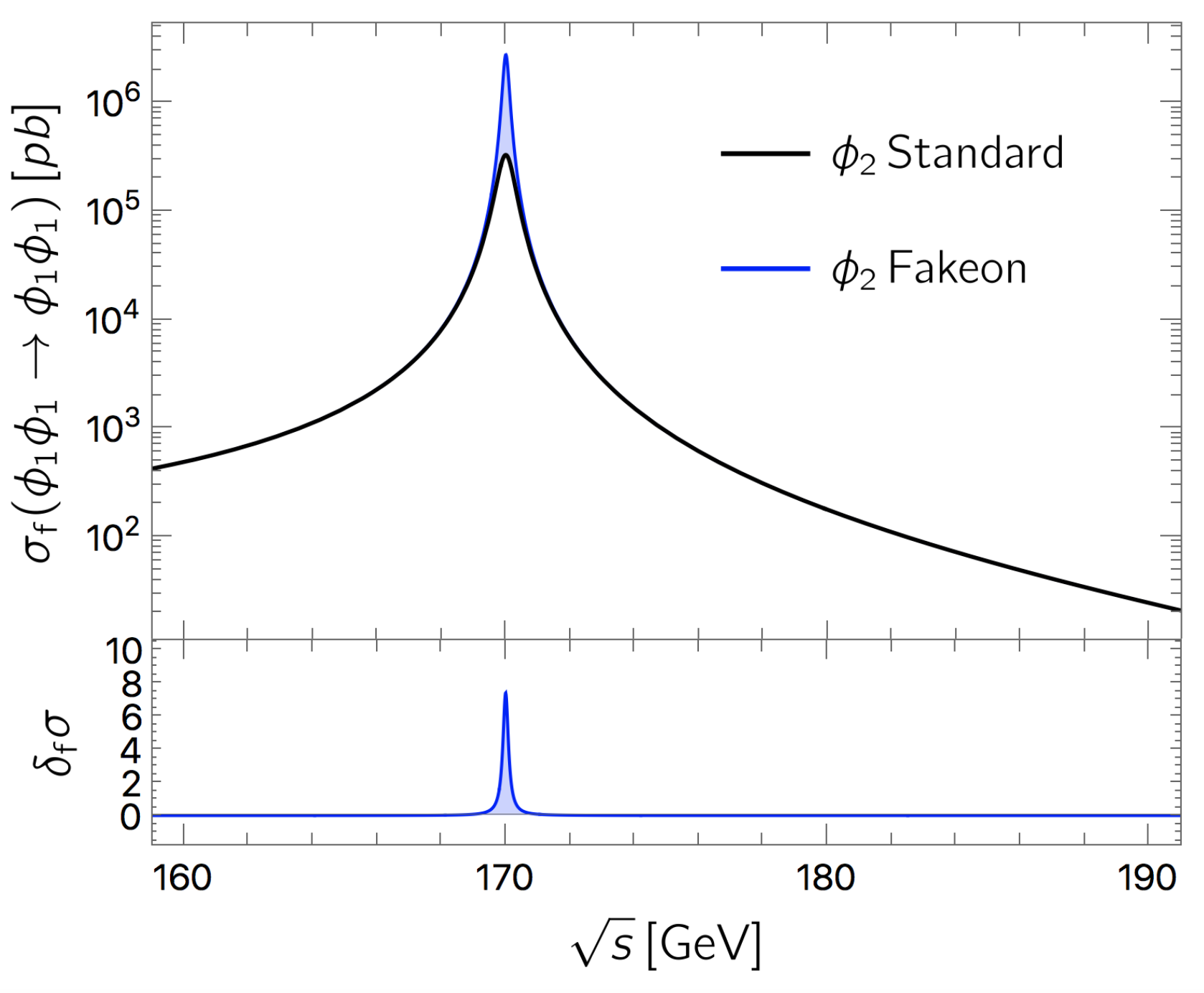}
    \vspace{-3mm}
    \caption{Top panel: cross section of the process $\phi_1\phi_1\rightarrow \phi_1\phi_1$ as a function of the center-of-mass energy $\sqrt{s}$ around the peak of $\phi_3$ in the case where $\phi_2$ is a standard particle (black line) or a fakeon (blue line). In this example the Lagrangian parameters are set to $(m_1,m_2,m_3,\lambda)=(10,80,170,60)$ GeV. Bottom panel: the quantity $\delta_{\rm f}\sigma$ as a function of $\sqrt{s}$. The modifications in the imaginary part of the diagrams in \autoref{fig:Gamma3_SelfEnergies} increase the height of the peak of almost one order of magnitude.}
    \label{fig:Sigma1111_Peak}
\end{figure}
 So no modifications arise from $\Delta_{\rm f} C_0$ and $\Delta_{\rm f} D_0$ at the one-loop level.
However, in the neighborhood of $\sqrt{s}=170$ GeV, the triangle and box diagrams are negligible with respect to the corrections coming from the bubble diagrams and we can resum the self energies. The dressed propagator reads
 \begin{equation}
     \frac{i}{s-m_3^2-\Sigma_{3} (s)},
 \end{equation}
 where $-i\Sigma_{3}$ is given by the sum of the self-energy diagrams of~\autoref{fig:Gamma3_SelfEnergies}. If $\phi_2$ is a fakeon, the second and third diagram in~\autoref{fig:Gamma3_SelfEnergies} have a vanishing imaginary part, since their cut diagrams are identically zero due to the fakeon prescription. This changes the width of $\phi_3$, which now can only decay into two $\phi_1$ particles. Choosing for simplicity $\lambda_{13}=\lambda_{23}=\lambda_{123}\equiv\lambda=60$ GeV, the widths in the two cases read
\begin{equation}
\label{eq:Gamma3}
    \Gamma_3=\Gamma_3^{11}+\Gamma_3^{22}+\Gamma_3^{12}\simeq 0.60 \ \text{GeV}, \qquad \Gamma_3^{\text{f}}=\Gamma_3^{11}\simeq 0.21 \ \text{GeV},
\end{equation}
\begin{equation}
    \Gamma_3^{ij}\equiv\frac{\lambda^2}{m_3}\text{Im}\left[B_0(m^2_3,m^2_i,m^2_j)\right]=\frac{\lambda^2}{16 \pi m_3^3}\sqrt{\lambda(m_3^2,m_i^2,m_j^2)},
\end{equation}
 where the superscripts $ij$ and ``f" denote the particles that circulate in the loop and the case where $\phi_2$ is a fakeon, respectively.
In the top panel of ~\autoref{fig:Sigma1111_Peak} we display the cross section $\sigma(\phi_1\phi_1\rightarrow \phi_1\phi_1)$ around $\sqrt{s}=m_3$, where $\phi_2$ is considered either as a standard particle (black line) or a fakeon (blue line).
The height of the peak is well approximated by $\sigma_{\rm (f)}(m^2_3)\simeq \lambda^4/(16 \pi\, m^4_3\,\Gamma^{\rm (f)2}_3)$ and using the values in \eqref{eq:Gamma3}, we find $\sigma(m^2_3)\simeq 3.3 \times 10^{5}$ pb in the standard case, while in the fakeon case $\sigma_{\rm f}(m^2_3)\simeq 2.7 \times 10^{6}$ pb, which amounts to a difference of almost an order of magnitude between the two scenarios as displayed in the bottom panel of~\autoref{fig:Sigma1111_Peak}. This gives a concrete example where the presence of a fakeon is experimentally detectable. We recall that most of the known particles are detected indirectly so we should question whether their fundamental nature is standard or purely virtual. In light of what we discuss in this paper, we should proceed with a one-by-one exclusion analysis based on the present experimental knowledge. For example, due to confinement, quarks are never directly detected. However the possibility for the light quarks to be fakeons is already ruled out by the precise measurement of the height of the Z peak at LEP. We stress that the difference would be only in a very narrow region surrounding the peak, similarly to what we show in~\autoref{fig:Sigma1111_Peak}. Another interesting example is the Higgs boson, which will be studied in detail in a dedicated paper~\cite{PivaMelis}.

Initial studies of physical consequences of these type of modifications have been done in the context of particle physics~\cite{Anselmi:2021icc,Anselmi:2021chp} and quantum gravity~\cite{Anselmi:2018tmf}.

\subsection{Modified cross section away from the peak}
 \begin{figure}[t]
    \centering
    \includegraphics[scale=0.33]{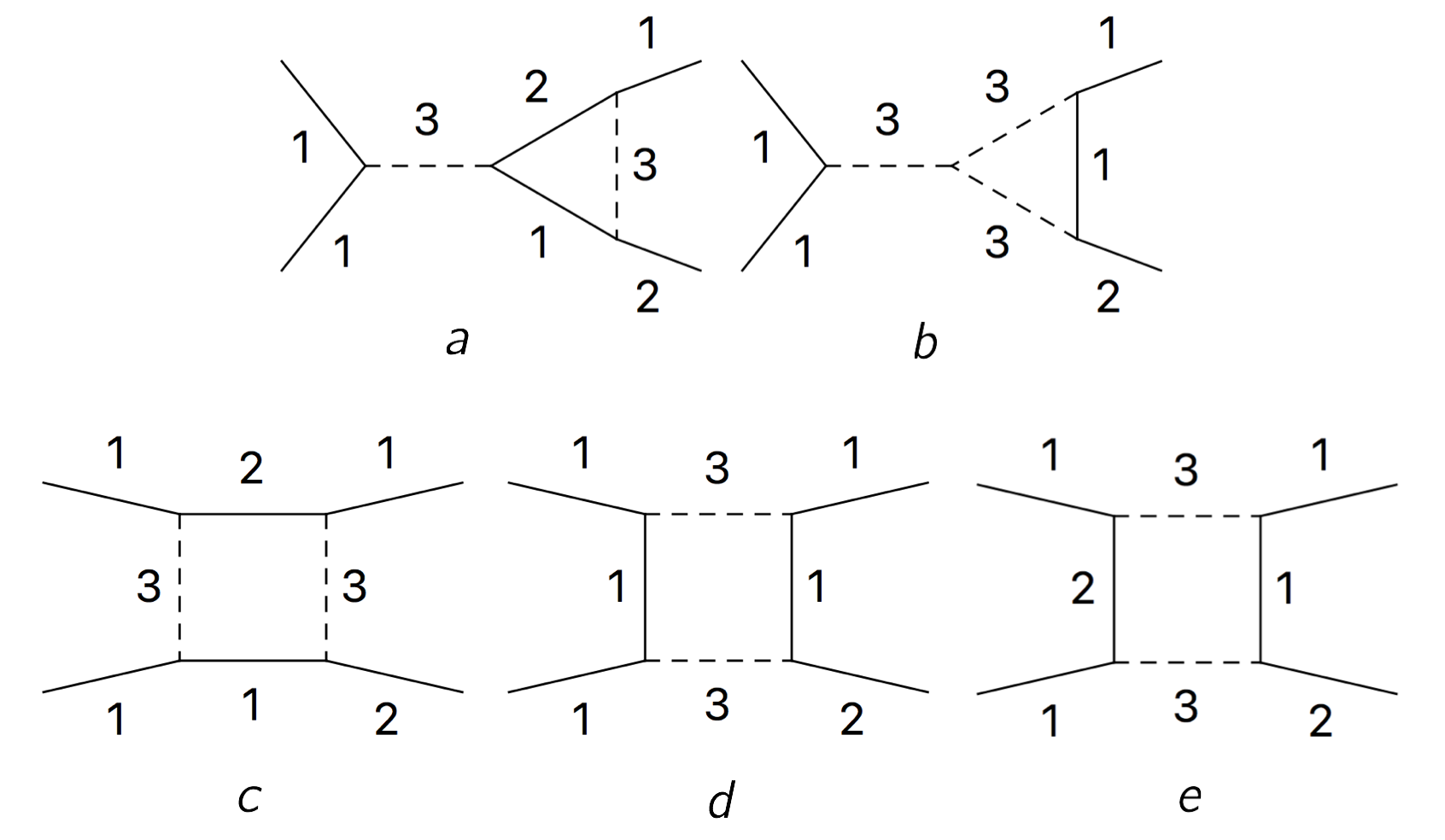}
    \captionsetup{format=hang}
    \caption{Triangle and box diagrams that develop nonvanishing $\Delta\Delta(\mathcal{Q})$-terms and modify the cross section of the process $\phi_1\phi_1\rightarrow \phi_1 \phi_2$. The dashed line identifies the fakeon.}
    \label{fig:Sigma1112_Diagrams}
\end{figure}
We consider a different scenario, where $\phi_3$ is a fakeon, and move to a region of $\sqrt{s}$ where the triangle and box diagrams cannot be neglected. In this case we assume a mass hierarchy such that $m_2 > m_1 + m_3$, so in the box and triangle diagrams at least one external momenta squared is always above threshold, while for another one it depends on $s$. Specifically, we set $(m_1, m_2, m_3)=(10, 170, 80)$ GeV and study the process $\phi_1\phi_1\rightarrow \phi_1\phi_2$. As discussed above, since only real modifications affect the cross section at this order, we are interested in the terms of the form~\eqref{eq:D12D13LorInv} and~\eqref{eq:DDQBox}.
\begin{figure}[t]
    \centering
    \includegraphics[scale=0.265]{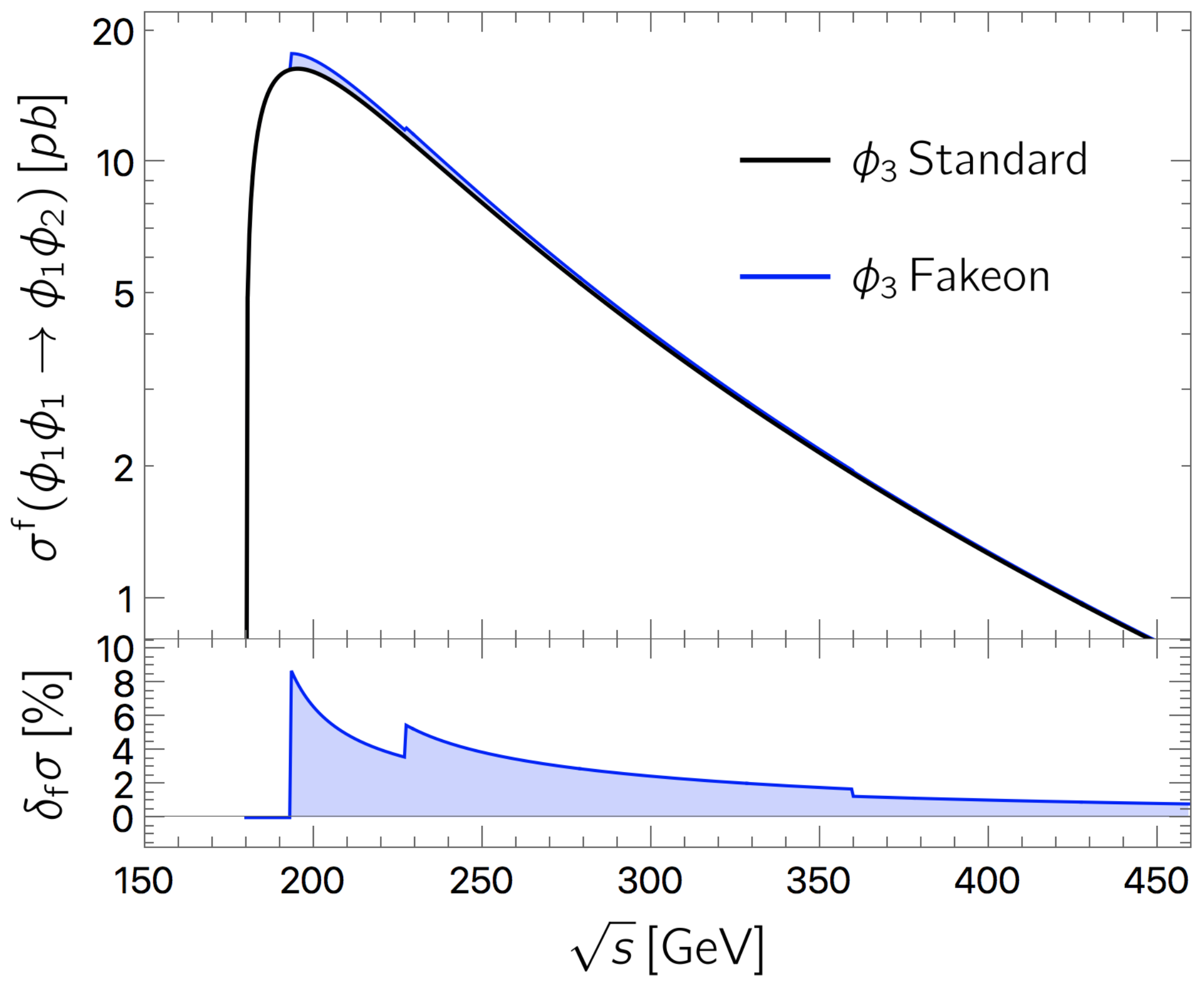}
    \caption{Top panel: cross section of the process $\phi_1\phi_1\rightarrow \phi_1\phi_2$ as a function of the center-of-mass energy $\sqrt{s}$, in the case where $\phi_3$ is a standard particle (black line) or a fakeon (blue line). The modifications in the real part of the diagrams in \autoref{fig:Sigma1112_Diagrams} are quantified through the quantity $\delta \sigma_{\rm f}$ displayed in the bottom panel. In this example the Lagrangian parameters are set to $(m_1,m_2,m_3,\lambda)=(10,170,80,60)$ GeV.}
    \label{fig:Sigma1112}
\end{figure}
In the case considered, the diagrams that have the required configurations to contribute to $\delta_{\text{f}}\sigma$ are shown in~\autoref{fig:Sigma1112_Diagrams}. Note that the triangle and the box diagrams (a,c) and (b,d,e) share the same conditions for nonvanishing $\Delta_{\rm f} C_0$ and $\Delta_{\rm f} D_0$. In particular, the TV-functions related to the diagrams (a,c) coincide with those studied in the examples of~\autoref{sec:PVmod} and displayed in~\autoref{fig:C0_example} and~\autoref{fig:D0_example}.
After integrating over the phase space and combining the conditions~\eqref{eq:condp2pm} and~\eqref{eq:DDQcond}, the diagrams (a,c) give a nonvanishing $\delta_{\text{f}}\sigma$ in the interval $227 \ \text{GeV}\lesssim \sqrt{s}\lesssim 360 \ \text{GeV}$, while the diagrams (b,d,e) contribute in a wider range $193 \ \text{GeV}\lesssim \sqrt{s}\lesssim 1192 \ \text{GeV}$. This is shown in the top panel of~\autoref{fig:Sigma1112}, where we can see the total cross section $\sigma_{\rm f}(\phi_1\phi_1\rightarrow \phi_1\phi_2)$ as a function of $\sqrt{s}$ in the case where $\phi_3$ is a standard particle (black line) or a fakeon (blue line). The corresponding behavior of $\delta_{\rm f} \sigma$ can be read from the bottom panel. The largest differences between the two scenarios are located at $\sqrt{s}=193$ GeV and $\sqrt{s}=227$ GeV, where $\delta_{\rm f} \sigma \simeq 9\% $ and $\delta_{\rm f} \sigma \simeq 6\% $, respectively. For $\sqrt{s}>227$ GeV the ratio $\delta_{\rm f} \sigma$ rapidly decreases, dropping below $1\% $ already at $\sqrt{s}=450$ GeV.
The nonanalyticities typical of these modifications are evident. We stress that this type of behavior is not unusual in a cross section. Indeed, discontinuities like those in~\autoref{fig:Sigma1112} can appear in a standard cross section at two loops, where the imaginary parts of the one-loop amplitudes must be taken into account. Furthermore, other nonanalyticities might also appear in a one-loop cross section, since the real parts of a one-loop amplitude typically have points where they are continuous but not differentiable. In standard cases, these effects are highly suppressed and often not visible. However, the fakeon prescription has the feature of turning those effects on (in regions where they are usually absent) and amplify them, as we have shown in~\autoref{sec:PVmod}. Moreover, the largest differences are always in the neighborhood of the points where the effects turn on. This is an important phenomenological feature, since it localizes the effects of this type of new physics. Indeed, it is possible to select out a relatively small range of center-of-mass energy where to expect such behavior and suggest targeted experiments.  
 
\vspace{-2mm} 

 \section{Conclusions}\label{sec:conclusions}
We have derived explicit expressions of one-loop scalar integrals in the case of purely virtual particles for the bubble, triangle and box diagrams, which appear in most applications in particle physics. The modified expressions are written in terms of subtractions from the usual TV-functions. The Passarino-Veltman reduction applies straightforwardly, which makes the results of this paper applicable to any quantum field theory.

The peculiarities of the fakeons prescription that defines purely virtual particles leads to modifications in both the real and imaginary parts of the amplitudes. In particular, new points of nonanaliticity appear, which are not associated to physical production of particles. However, such a feature can be useful from the phenomenological point of view, since it allows us to discriminate models with fakeons from those without. We have given a concrete example by means of a toy model, where we study different scenarios in which differences between standard particles and fakeons can be directly observed in cross sections.

Moreover, fakeon phenomenology can depart from standard phenomenology even more radically. In this sense it is important to note that fakeons could induce effective long-range interactions at the one loop level. The new-physics effects offered by this possibility are worth of specific investigations.

The modified TV-functions can be easily implemented in software such as {\tt LoopTools} and {\tt FormCalc} and used to compute large amount of one-loop Feynman diagrams in theories with fakeons. The main application is the study of fakeons in the SM and BSM models. Our rationale is that, since quantum field theory allows fakeons, everytime the existence of a particle is inferred from indirect detection, further tests should be carried out in order to establish whether such a particle is a fakeon or a standard one. Finally, in the same fashion several BSM models can be (re)formulated, in view of the fact that many experimental constraints that rule out the most conventional models are eluded in the case of fakeons. The functions provided in this paper are a fundamental ingredient to pave the way to such kind of analysis.
Indeed, we believe that the concept of purely virtual particles opens unexplored scenarios in particle physics phenomenology, where all the experimental searches so far have not provided convincing evidence for the existence of physics beyond the SM.

\subsection*{Acknowledgments}
We thank D. Anselmi, C. Marzo and L. Marzola for useful discussions. This work was supported by the Estonian Research Council grant MOBTT86.

\appendix
\renewcommand{\thesection}{\Alph{section}} \renewcommand{\theequation}{%
\thesection.\arabic{equation}} \setcounter{section}{0}

\sect{Computation details of one-loop integrals}\label{appcomp}
In this appendix we show explicitly the steps for the computation of the relevant integrals used in this paper. We label the internal and external momenta according to the {\tt LoopTools} conventions in Figure \ref{fig:Mom_Conv}. We adopt spherical coordinates $(q_s, \theta, \varphi)$ and, to simplify the notation, we rename $(\cos\theta,\cos\varphi)=(u,v)$. For the examples displayed, we work in the frame where $\boldsymbol{p_1}=0$. For each result we show the substitutions used to recover Lorentz invariance. The signs of the energies $p_i^0$ are chosen consistently with the arguments of the $\delta$-functions in the integrals.

We start with the simplest case of the bubble diagram and compute the term $\Delta^{12}$
\begin{eqnarray}
\label{eq:D12comp}
\int\frac{\mathrm{d}^3\boldsymbol{q}}{(2 \pi)^3}\frac{ \Delta^{12}}{4\,\omega_1\omega_2}&=&\int^{\infty}_{0}\frac{\mathrm{d} q_s \, q_s^2}{2\pi^2}\frac{ \pi\,\delta(p_1^0-\omega_1-\omega_2)}{4\,\omega_1\omega_2} = \frac{\theta(p^0_1-m_1-m_2)\,q_{s12}}{8\pi p^0_1} \,,
\end{eqnarray}
where $\omega_1=\sqrt{q_s^2+m_1^2}$, $\omega_2=\sqrt{q_s^2+m_2^2}$ and
 \begin{equation}
\label{eq:qs0}
q_{s12} =\sqrt{\frac{\lambda((p^{0}_{1})^2,m^2_1,m^2_2)}{4(p^{0}_{1})^2}}
 \end{equation}
is the zero of the argument of the $\Delta^{12}$. The Lorentz invariant form of \eqref{eq:D12comp} is recovered with the substitution $p^0_1=\sqrt{p^2_1}$, this gives the result in \eqref{eq:D12term} of Section \ref{sec:B0f_function}.

In the case of the triangle diagram we compute the integrals of $\Delta^{12}\mathcal{Q}^{13}$ and $\Delta^{12}\Delta^{13}$ since all the others are derived by means of cyclic permutations of the indices $(1,2,3)$. We have $\omega_1=\sqrt{q_s^2+m_1^2}$, $\omega_2=\sqrt{q_s^2+m_2^2}$, $\omega_3=\sqrt{q_s^2+p^2_{s3}-2 q_s p_{s3} u+m_3^2}$. 

The integral of the $\Delta\mathcal{Q}$-type is given by
\begin{eqnarray}
\label{eq:DQcomp}
\int\frac{\mathrm{d}^3\boldsymbol{q}}{(2 \pi)^3}\frac{ \Delta^{12}\mathcal{Q}^{13}}{8\,\omega_1\omega_2\omega_3}
&=&\int^{\infty}_{0}\frac{\mathrm{d} q_s \, q_s^2}{(2 \pi)^2}\int^{1}_{-1} \hspace{-2mm}\mathrm{d}u\frac{\pi \delta(p_1^0-\omega_1-\omega_2)}{8\,\omega_1\omega_2\omega_3}\mathcal{P}\frac{2 \omega_3}{(p_3^0+\omega_1)^2-\omega_3^2}\nonumber\\[7pt]
&=&\frac{\theta(p^0_1-m_1-m_2)\,q_{s12}}{16\pi p^0_1}\int^{1}_{-1} \hspace{-2mm}\mathrm{d}u\,\mathcal{P}\frac{1}{(p_3^0+\overline{\omega}_1)^2-\overline{\omega}_3^2}\\[7pt]
&=& \frac{\theta(p^0_1-m_1-m_2)}{32\pi\,p^0_1 \, p_{s3}}\,{\rm ln}\left|\frac{u_{13}+1}{u_{13}-1}\right|\,,\nonumber
\end{eqnarray}
where in the second step of \eqref{eq:DQcomp} we used the result of \eqref{eq:D12comp} to perform the integration in $q_s$, and
\begin{eqnarray}
\label{eq:ctheta}
&& u_{13}  = \frac{(p_3^0)^2-p^2_{s3}+2\,p_3^0\sqrt{q^2_{s12}+m^2_1}+m^2_1-m^2_3}{2\, p_{s3}\,q_{s12}}
\end{eqnarray}
is the zero of the argument of the $\Delta^{13}$.

The integral of the $\Delta\Delta$-type is
\begin{eqnarray}
\label{eq:DDcomp}
\int\frac{\mathrm{d}^3\boldsymbol{q}}{(2 \pi)^3}\frac{ \Delta^{12}\Delta^{13}}{8\,\omega_1\omega_2\omega_3}
&=&\int^{\infty}_{0}\frac{\mathrm{d} q_s \, q_s^2}{(2 \pi)^2}\int^{1}_{-1} \hspace{-2mm}\mathrm{d}u\frac{\pi^2 \delta(p_1^0-\omega_1-\omega_2)\delta(p_3^0+\omega_1+\omega_3)}{8\,\omega_1\omega_2\omega_3} \nonumber\\[7pt]
&=&\frac{\theta(p^0_1-m_1-m_2)\,q_{s12}}{32\,p^0_1}\int^{1}_{-1} \hspace{-2mm}\mathrm{d}u \frac{\delta(p^0_3+\overline{\omega}_1+\overline{\omega}_3)}{\overline{\omega}_3} \\[7pt]
&=&\frac{\theta\left(p^0_1-m_1-m_2,m_2^2-m_1^2-p_1^0\left(p_1^0+2p_3^0\right),1-|u_{13}|\right)}{32\,p^0_1 \, p_{s3}} \nonumber\,,
\end{eqnarray}
where, by definition, $u_{13}$ is given by \eqref{eq:ctheta}. In order to recast the expressions above in an explicit Lorentz invariant form we apply the substitutions
\begin{eqnarray}
& p^0_1=\sqrt{p^2_1},\quad p^0_3=-\cfrac{p^2_3-p^2_{2}+p^2_1}{2\sqrt{p^2_1}},\quad p_{s3}=\sqrt{\cfrac{\lambda(p^2_1,p^2_2,p^2_{3})}{4p^2_1}}\,.\hspace{4mm}
\end{eqnarray}
The final results are given in \eqref{eq:D12Q13LorInv} and \eqref{eq:D12D13LorInv} of Section \ref{sec:C0f_function}.

In the case of the box diagram we compute the integrals of $\Delta^{12}\mathcal{Q}^{13}\mathcal{Q}^{14}$, $\Delta^{12}\Delta^{13}\mathcal{Q}^{14}$ and $\Delta^{12}\Delta^{13}\Delta^{14}$, all the others are derived by means of permutations of the indices $(1,2,3,4)$. We have $\omega_1=\sqrt{q_s^2+m_1^2}$, $\omega_2=\sqrt{q_s^2+m_2^2}$, $\omega_3=\sqrt{q_s^2+p^2_{s2}+2 q_s p_{s2}u+m_3^2}$ and $\omega_4=\sqrt{q_s^2+p^2_{s4}-2 q_s p_{s4}(s_{24} \sqrt{1-u^2}v + c_{24} u)+m_4^2}$, $s_{24}$ and $c_{24}$ being the sine and cosine of the angle between $\boldsymbol{p}_2$ and $\boldsymbol{p}_4$, respectively.

The integral of the $\Delta\mathcal{Q}\mathcal{Q}$-type is
\begin{eqnarray}
\label{eq:DQQexp}
\int&& \hspace{-4mm}\frac{\mathrm{d}^3\boldsymbol{q}}{(2 \pi)^3} \frac{ \Delta^{12}\mathcal{Q}^{13}\mathcal{Q}^{14}}{16\,\omega_1\omega_2\omega_3\omega_4}
=\nonumber\\[7pt]
&=&\int^{\infty}_{0}\frac{\mathrm{d} q_s \, q_s^2}{(2 \pi)^3}\int^{1}_{-1} \hspace{-2mm}\mathrm{d}u\int^{2\pi}_{0} \hspace{-3mm}\mathrm{d}\varphi \frac{\pi \delta(p_1^0-\omega_1-\omega_2)}{16\,\omega_1\omega_2\omega_3\omega_4} \mathcal{P}\frac{2 \omega_3}{(p_1^0-\omega_1)^2-\omega_3^2}\mathcal{P}\frac{2 \omega_4}{(p_4^0+\omega_1)^2-\omega_4^2}\nonumber\\[7pt]
&=&\frac{\theta(p^0_1-m_1-m_2)}{32\pi^2\,p^0_1} \int^{1}_{-1} \hspace{-2mm}\mathrm{d}u \int^{2\pi}_{0} \hspace{-3mm}\mathrm{d}\varphi \,\mathcal{P}\frac{1}{(p_1^0-\overline{\omega}_1)^2-\overline{\omega}_3^2}\,\mathcal{P}\frac{1}{(p_4^0+\overline{\omega}_1)^2-\overline{\omega}_4^2}\\[9pt]
&=&  \frac{\theta(p^0_1-m_1-m_2)}{32\pi\, p^0_1\,p_{s2}\,p_{s4}\,q_{s12}} \int^{1}_{-1} \hspace{-2mm}\mathrm{d}u\, \mathcal{P}\frac{\text{sgn}(u_{14}+c_{24}u) \ \theta\left(\left|\frac{u_{14}+c_{24}u}{s_{24}\sqrt{1-u^2}}\right|-1\right)}{ (u_{13}-u)\sqrt{(1-c^2_{24})(1-u^2)\left[\left(\frac{u_{14}+c_{24}u}{s_{24}\sqrt{1-u^2}}\right)^2-1\right]}}\hspace{5mm} \nonumber\\
&=& \frac{\theta(p^0_1-m_1-m_2)}{32\pi\,p^0_1\,p_{s2}\,p_{s4}\,q_{s12}\sqrt{(1-c^2_{24})(1-u_{13}^2)|v_{14}^2-1|}} \nonumber\\[9pt]
&\times&
\begin{cases}
 \text{sgn}(u_{14}+c_{24}u)\,\text{ln}\left|\sdfrac{\kappa - (u_{13}+u_{14}c_{24})(u_{13}-u) +|u_{14}+c_{24}u|\sqrt{\kappa}}{(u_{13}-u)\sqrt{(1-u^2_{14})(1-c^2_{24})}}\right|^{u=1}_{u=-1}  & \text{if $|v_{14}|>1$ }\\[15pt]
\text{sgn}(u_{14}+c_{24}u)\,\text{arcsin}\left[\sdfrac{\kappa- (u_{13}+u_{14}c_{24})(u_{13}-u)}{ |u_{13}-u|\sqrt{(1-u^2_{14})(1-c^2_{24})}} \right]^{u=1}_{u=u_+}  & \text{if $|v_{14}|<1$ } \\[10pt]
+\, \text{sgn}(u_{14}+c_{24}u)\,\text{arcsin}\left[\sdfrac{\kappa- (u_{13}+u_{14}c_{24})(u_{13}-u)}{ |u_{13}-u|\sqrt{(1-u^2_{14})(1-c^2_{24})}} \right]^{u=u_-}_{u=-1}   \quad ,
\end{cases}\hspace{0.7cm}   
\end{eqnarray}
where $\kappa \equiv (1-c^2_{24})(1-u^2_{13})(v_{14}^2-1)=u^2_{13}+u^2_{14}+c^2_{24}+2u_{13}u_{14}c_{24} -1$, symmetric under the exchange of its arguments, $u_{\pm} \equiv -u_{14}c_{24}\pm \sqrt{(1-u^2_{14})(1-c^2_{24})}$ are the zeros of the argument of the $\theta$-function in the fourth line of \eqref{eq:DQQexp}, and $v_{14}$, $u_{13}$, $u_{14}$, $c_{24}$ are 
\begin{eqnarray}
\label{eq:cphi14}
v_{14} &=&   \frac{u_{14}+ c_{24}u_{13} }{\sqrt{(1-c^2_{24})(1-u^2_{13})}}\, , \quad c_{24} = \frac{p^0_2\,p^0_4-p_2\cdot p_4}{p_{s2}\,p_{s4}} \nonumber \\[7pt]
 u_{13} &=&\frac{(p_{12}^0)^2-p^2_{s2}-2\,p_{12}^0\sqrt{q^2_{s12}+m^2_1}+m^2_1-m^2_3}{2\,p_{s2}\,q_{s12}}\,,\\
 u_{14} &=&\frac{(p_{4}^0)^2-p^2_{s4}+2\,p_{4}^0\sqrt{q^2_{s12}+m^2_1}+m^2_1-m^2_4}{2\,p_{s4}\,q_{s12}}\,, \nonumber
\end{eqnarray}
where $u_{13}$ and $v_{14}$ are the zeros of the arguments of the $\Delta^{13}$ and $\Delta^{14}$, respectively.

The integral of the $\Delta\Delta\mathcal{Q}$-type is
\begin{eqnarray}
\int && \hspace{-5mm}\frac{\mathrm{d}^3\boldsymbol{q}}{(2 \pi)^3} \frac{ \Delta^{12}\Delta^{13}\mathcal{Q}^{14}}{16\,\omega_1\omega_2\omega_3\omega_4}
=\nonumber\\[7pt]
&=&\int^{\infty}_{0}\frac{\mathrm{d} q_s \, q_s^2}{(2 \pi)^3}\int^{1}_{-1} \hspace{-2mm}\mathrm{d}u\int^{2\pi}_{0} \hspace{-3mm}\mathrm{d}\varphi \frac{\pi^2\delta(p_1^0-\omega_1-\omega_2)\delta(p_{12}^0-\omega_1-\omega_3)}{16\,\omega_1\omega_2\omega_3\omega_4} \mathcal{P}\frac{2 \omega_4}{(p_4^0+\omega_1)^2-\omega_4^2}\nonumber\\[7pt]
&=&\frac{\theta(p^0_1-m_1-m_2,m_2^2-m_1^2-p_1^0\left(p_1^0+2p_{12}^0\right),1-|u_{13}|)}{64\pi\,p^0_1\,p_{s2}}\int^{2\pi}_{0} \hspace{-3mm}\mathrm{d}\varphi \,\mathcal{P}\frac{1}{(p_4^0+\overline{\omega}_1)^2-\overline{\omega}_4^2} \hspace{5mm}\\[7pt]
&=&\frac{\text{sgn}(v_{14})\theta(p^0_1-m_1-m_2,m_2^2-m_1^2-p_1^0\left(p_1^0+2p_{13}^0\right),1-|u_{13}|,|v_{14}|-1)}{64\,p^0_1\,p_{s2}\,p_{s4}\,q_{s12} \sqrt{(1-c^2_{24})(1-u^2_{13})(v_{14}^2-1)}}\,. \nonumber
\end{eqnarray}

The integral of the $\Delta\Delta\Delta$-type is
\begin{eqnarray}
\int && \hspace{-5mm}\frac{\mathrm{d}^3\boldsymbol{q}}{(2 \pi)^3} \frac{ \Delta^{12}\Delta^{13}\Delta^{14}}{16\,\omega_1\omega_2\omega_3\omega_4}
=\nonumber\\[7pt]
&=&\int^{\infty}_{0}\frac{\mathrm{d} q_s \, q_s^2}{(2 \pi)^3}\int^{1}_{-1} \hspace{-2mm}\mathrm{d}u\int^{2\pi}_{0} \hspace{-3mm}\mathrm{d}\varphi \frac{\pi^3 \delta(p_1^0-\omega_1-\omega_2)\delta(p_{12}^0-\omega_1-\omega_3)\delta(p_4^0+\omega_1+\omega_4)}{16\,\omega_1\omega_2\omega_3\omega_4} \nonumber\\[7pt]
&=&\frac{\theta(p^0_1-m_1-m_2,m_2^2-m_1^2-p_1^0\left(p_1^0+2p_{12}^0\right),1-|u_{13}|)}{128\,p^0_1\,p_{s2}}\int^{2\pi}_{0} \hspace{-3mm}\mathrm{d}\varphi \,\frac{\delta(p_4^0+\overline{\omega}_1+\overline{\omega}_4)}{\overline{\omega}_4} \hspace{5mm}\\[7pt]
&=&\frac{\theta(p^0_1-m_1-m_2,m_2^2-m_1^2-p_1^0\left(p_1^0+2p_{12}^0\right),1-|u_{13}|,m_2^2-m_1^2-p_1^0\left(p_1^0-2p_{4}^0\right),1-|v_{14}|)}{64\,\,p^0_1\,p_{s2}\,p_{s4}\,q_{s12} \sqrt{(1-c^2_{24})(1-u^2_{13})(1-v_{14}^2)}}\,. \nonumber
\end{eqnarray}
All the expressions above are recasted into a Lorentz invariant form through the substitutions 
\begin{eqnarray}
&p^0_1=\sqrt{p^2_1},\quad p^0_{12}=\cfrac{p^2_{12}-p^2_2+p^2_1}{2\sqrt{p^2_1}},\quad p^0_4=-\cfrac{p^2_4-p^2_{23}+p^2_1}{2\sqrt{p^2_1}},\quad q_{s12}=\sqrt{\cfrac{\lambda(p^2_1,m^2_1,m^2_2)}{4p^2_1}}\,, \nonumber\\
&p_{s2}=\sqrt{\cfrac{\lambda(p^2_1,p^2_2,p^2_{12})}{4p^2_1}},\quad p_{s4} =\sqrt{\cfrac{\lambda(p^2_1,p^2_4,p^2_{23})}{4p^2_1}}, \quad p_2 \cdot p_4 =\cfrac{p^2_1+p^2_3-p^2_{12}-p^2_{23}}{2}.
\end{eqnarray}
The Lorentz invariant results are given in \eqref{eq:DQQBox}, \eqref{eq:DDQBox} and \eqref{eq:DDDBox1} of Section \ref{sec:D0f_function}.

\bibliographystyle{JHEP} 
\bibliography{mybiblio}

\end{document}